\pdfoutput=1
\documentclass{article}
\usepackage{arxiv}

\usepackage[utf8]{inputenc}
\usepackage[T1]{fontenc}
\usepackage{amsmath,amssymb,bm}
\usepackage{graphicx}
\usepackage{booktabs}
\usepackage{array}
\usepackage[round,authoryear]{natbib}
\usepackage{xcolor}
\usepackage{url}
\usepackage{etoolbox}
\usepackage{microtype}
\usepackage{placeins}
\usepackage{hyperref}
\hypersetup{hidelinks}
\graphicspath{{figures/}}
\newcommand{\Retau}{\ensuremath{\mathrm{Re}_{\tau}}}
\newcommand{\nl}{\ensuremath{n_{\ell}}}
\newcommand{\dz}{\ensuremath{d_z}}
\newcommand{\ks}{\ensuremath{k_{\mathrm{s}}}}
\newcommand{\kd}{\ensuremath{k_{\mathrm{d}}}}

\renewcommand{\shorttitle}{Effective-dimension estimation of a minimal turbulent channel flow via IRMAE-WD}
\renewcommand{\headeright}{}
\renewcommand{\undertitle}{}
\makeatletter
\patchcmd{\@maketitle}{\textsc{\undertitle}\\}{}{}{\PackageWarning{preamble}{undertitle line not removed}}
\makeatother
\hypersetup{pdftitle={Effective-dimension estimation of a minimal turbulent channel flow via implicit rank-minimizing autoencoders},
  pdfauthor={Ryuto Saeki, Keita Nakamura, Koji Fukagata}}

\newcommand{\orcid}[2]{\href{https://orcid.org/#1}{\includegraphics[scale=0.06]{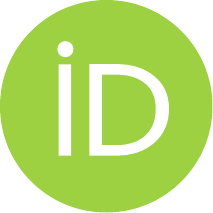}\hspace{1mm}#2}}
\title{Effective-dimension estimation\\ of a minimal turbulent channel flow\\ via implicit rank-minimizing autoencoders}
\author{\orcid{0009-0003-0532-0732}{Ryuto Saeki}\\
  Department of Mechanical Engineering\\
  Keio University\\
  Yokohama 223-8522, Japan\\
  \texttt{saekiryuto@keio.jp}
  \And \orcid{0009-0009-7098-2269}{Keita Nakamura}\\
  Department of Mechanical Engineering\\
  Keio University\\
  Yokohama 223-8522, Japan\\
  \texttt{keita.n\_0925@keio.jp}
  \And \orcid{0000-0003-4805-238X}{Koji Fukagata}\\
  Department of Mechanical Engineering\\
  Keio University\\
  Yokohama 223-8522, Japan\\
  \texttt{fukagata@mech.keio.ac.jp}}
\date{\today}

\begin{document}
\maketitle

\begin{abstract}
The implicit rank-minimizing autoencoder with weight decay (IRMAE-WD) (Zeng, K., P{\'e}rez De Jes{\'u}s, C. E., Fox, A. J. \& Graham, M. D. 2024 Mach. Learn.: Sci. Technol. 5, 025053) aims to estimate the dimension of the low-dimensional manifold underlying high-dimensional data from a sharp drop in the latent singular values of a single trained model.
We apply IRMAE-WD with a three-dimensional convolutional autoencoder to velocity fields of a minimal turbulent channel flow at $\Retau=110$ and examine whether the drop represents the intrinsic degrees of freedom.
We also compare the number of coordinates IRMAE-WD needs for a given reconstruction accuracy or smallest reproduced scale with that of proper orthogonal decomposition (POD).
We find that a clear drop appears and that the reconstruction rate decreases by at most 0.01 percentage points with truncation at the drop.
The drop thus marks the effective dimension of the trained representation.
However, the drop position moves to the lower-dimension side as linear layers are added.
In addition, the reconstruction error decreases only gradually with the number of coordinates in all but the smallest wavelength band.
These results suggest that the drop position is set by the implicit regularization of the linear layers and cannot be regarded as the intrinsic degrees of freedom.
In contrast, with transfer learning, the smallest number of coordinates needed for a given accuracy hardly depends on the number of linear layers and is smaller than that of POD up to a reconstruction rate of about 97\%.
IRMAE-WD also reproduces a given scale with fewer coordinates than POD.
Above about 97\%, however, POD requires fewer coordinates, mainly because IRMAE-WD reconstructs large-scale structures less precisely.
Hence, when the error decreases continuously, as in the minimal channel flow, the number of coordinates needed to represent wall turbulence should be stated with the required accuracy or scale, not as a single number.
\end{abstract}

\section{Introduction}
\label{sec:introduction}

Turbulence governs the skin-friction drag and the transport of heat and mass in pipes, aircraft, and heat exchangers.
Because wall turbulence is closely linked to the generation of skin-friction drag, understanding its structure and building models for its prediction and control are important in engineering \citep{Jimenez2018Structures}.
Direct numerical simulation (DNS), which resolves all scales of turbulent motion without a turbulence model, provides detailed velocity fields of turbulence and is a standard tool for analyzing flow phenomena.
However, resolving the wide range of spatial scales in turbulence requires a large number of grid points \citep{MoinMahesh1998}, and the flow state becomes very high-dimensional.
Such high-dimensional data are difficult to use directly for the prediction and control of the flow.
A low-dimensional description of the flow state with a small number of variables is thus desired \citep{Taira2017Overview,BruntonNoackKoumoutsakos2020}.

Proper orthogonal decomposition (POD) is a linear method to describe the flow state with a few coordinates \citep{Sirovich1987,Berkooz1993POD}.
For a prescribed number of modes, it gives the linear subspace that minimizes the reconstruction error of the given data \citep{Berkooz1993POD}.
Owing to this property, POD has been widely used for the analysis and modeling of flows \citep{Taira2020Applications}, including the extraction of coherent structures in turbulent channel flow \citep{MoinMoser1989} and low-dimensional models of cylinder wakes \citep{Noack2003}.
However, a POD reconstruction is restricted to linear combinations of the modes.
As a result, when the flow states lie on a nonlinear manifold, the linear subspace that contains the manifold may require many more dimensions than the manifold itself, and a small number of POD modes may not give sufficient accuracy \citep{LeeCarlberg2020,Peherstorfer2022}.
The number of POD modes thus gives only an upper bound on the intrinsic degrees of freedom, i.e., the minimum number of independent variables needed to describe the flow.
This has motivated methods that achieve the same accuracy with fewer coordinates by using nonlinear mappings.

The autoencoder (AE) is a representative method for nonlinear dimensionality reduction \citep{Kramer1991,HintonSalakhutdinov2006,VinuesaBrunton2022}.
An AE is an unsupervised (or self-supervised) learning method that trains an encoder and a decoder simultaneously to minimize the reconstruction error: the encoder compresses the input into low-dimensional latent variables, and the decoder reconstructs the input from them.
Owing to the nonlinear activation functions in the neural networks, an AE can represent a nonlinear mapping between the input and the output.
In fluid dynamics, \citet{MilanoKoumoutsakos2002} were, to the best of our knowledge, the first to apply an AE to turbulent flow fields and compare it with POD.
They showed that an AE with linear activation functions is equivalent to POD and that a nonlinear AE reconstructs the near-wall velocity fields of turbulent channel flow better than POD with the same number of latent variables.
In recent years, AEs based on convolutional neural networks (CNNs) \citep{LeCun1998}, referred to as CNN-AEs, have been developed as an efficient way to handle the spatial structures of flow fields \citep{Brenner2019Perspective,LeeCarlberg2020,Vinuesa2024,FukagataFukami2025,Fukami2026Guide}.
These CNN-AEs have been used for the nonlinear mode decomposition of cylinder wakes \citep{Murata2020Modes,Fukami2020Hierarchical} and turbulent flows \citep{Eivazi2022}, for the reduced-order modeling of flows around bluff bodies \citep{Hasegawa2020Reynolds,Hasegawa2020Bluff}, and for low-dimensional representations of two-dimensional turbulence \citep{PageBrennerKerswell2021}.
They have also been applied to the reduced-order modeling of minimal turbulent channel flows \citep{Nakamura2021Channel,Racca2023}.
In an AE, however, the width of the latent variables has to be specified before training.
When models with different widths are trained, the reconstruction error tends to increase as the width decreases \citep{Fukami2021ModelReduction,Racca2023}, and small-scale structures are no longer captured \citep{Nakamura2021Channel}.
Because the error changes continuously with the width, it is difficult to uniquely determine the number of coordinates needed to represent the flow.

To address this problem, AE-based methods that estimate the required number of coordinates from data have been proposed.
\citet{LinotGraham2020,LinotGraham2022} assumed that the long-time behavior of a dissipative system lies near a lower-dimensional invariant manifold in the state space \citep{FloryanGraham2022}, and estimated the dimension of this manifold by training AEs with different latent widths.
For the Kuramoto--Sivashinsky equation in several domain sizes, they took the width beyond which the reconstruction error no longer decreases sharply as the dimension.
A similar approach has been applied to two-dimensional Kolmogorov flow \citep{PerezDeJesusGraham2023}.
These approaches, however, require a model to be retrained for each latent width.
In contrast, the implicit rank-minimizing autoencoder with weight decay (IRMAE-WD) proposed by \citet{Zeng2024IRMAEWD} provides both the manifold dimension and the coordinates from a single trained model.
It estimates the dimension from the number of directions that the latent variables actually use, rather than from the specified latent width.
IRMAE-WD extends the implicit rank-minimizing autoencoder (IRMAE) of \citet{Jing2020IRMAE}, which stacks linear layers without activation functions between the encoder and the decoder.
Because gradient-based training of a product of matrices tends to select low-rank solutions \citep{Arora2019DeepMF,Sun2021GreedyRank}, a property referred to as implicit regularization, the variations of the latent variables concentrate in fewer directions than the specified width.
With linear layers alone, however, the singular values in the unnecessary directions do not decrease steadily during training, and the singular values drop much less sharply than when weight decay is added \citep{Zeng2024IRMAEWD}.
The weight decay in IRMAE-WD, which shrinks the weights toward zero at every update, resolves these problems, and the number of directions used is given by the position at which the singular values drop sharply, hereafter referred to as the drop position \citep{Zeng2024IRMAEWD}.
Since then, IRMAE-WD has been used for a low-dimensional model of two-dimensional Kolmogorov flow \citep{PerezDeJesus2025Symmetry}. For the same flow, it has also been used to analyze the Reynolds-number dependence of the attractor dimension, but the obtained rank changes with training conditions such as the mini-batch size and is treated as an upper bound on the dimension \citep{ClearyPage2025}.
For two-dimensional Rayleigh--B\'enard convection, a multi-scale flow, it has been reported that the required dimension is difficult to judge from the latent singular values alone \citep{Vinograd2025RBC}.
The meaning of the drop in the singular values thus needs to be examined for each flow of interest.
Moreover, these applications are limited to two-dimensional flows, and applications of dimension-estimating AEs to wall turbulence are still limited \citep{LinotGraham2023Couette}.
For the three-dimensional velocity fields of wall turbulence, it remains to be examined whether the drop position depends on the network configuration and the training, and whether the representation truncated at the drop position needs fewer coordinates than POD for the same accuracy.

In the present study, we apply IRMAE-WD based on a three-dimensional CNN-AE directly to the velocity fields of a minimal turbulent channel flow at the friction Reynolds number of $\Retau=110$, and we examine the relationship between the singular value spectrum of the latent variables, hereafter referred to as the latent spectrum, and the intrinsic degrees of freedom of the flow.
First, we vary the number of linear layers, the latent width, and the training progress, and examine how the shape of the latent spectrum and the drop position depend on these settings.
Second, we keep only the first $k$ directions of the latent variables in descending order of the singular values, and reconstruct the velocity fields from these $k$ coordinates.
By comparing the result with POD and with a CNN-AE without linear layers, we evaluate the trade-off between the reconstruction accuracy and the number of coordinates.
Third, we evaluate the turbulence statistics of the truncated reconstructions and identify the information on turbulence that is retained or lost when the number of coordinates is reduced.
Finally, we pair the required number of coordinates with the spatial scales that can be reproduced, and discuss how the number of coordinates needed to represent wall turbulence should be stated.

The rest of this paper is organized as follows.
Section~\ref{sec:method} describes the DNS and the datasets, the three-dimensional CNN-AE and IRMAE-WD, the two training procedures (training from random initialization and transfer learning), the singular value decomposition and truncation of the latent variables, the methods compared with IRMAE-WD (a baseline CNN-AE and POD), and the evaluation metrics.
Section~\ref{sec:results} presents and discusses, in turn, the drop position and its meaning for the models trained from random initialization, the latent spectra and the number of coordinates for the models obtained by transfer learning, the statistics of the reconstructed fields, and the relationship between the reproducible scales and the required number of coordinates.
Finally, conclusions are drawn in section~\ref{sec:conclusion}.
\section{Methodology}
\label{sec:method}

\subsection{Direct numerical simulation}
\label{subsec:method_dns}

We consider a minimal turbulent channel flow driven by a constant pressure gradient.
The governing equations are the incompressible continuity and Navier--Stokes equations,
\begin{align}
 &\bm\nabla\cdot\bm u=0,
 \label{eq:method_continuity}\\
 &\frac{\partial\bm u}{\partial t}
 =-\bm\nabla\cdot(\bm u\bm u)-\bm\nabla p
   +\frac{1}{\Retau}\nabla^2\bm u,
 \label{eq:method_ns}
\end{align}
where $\bm u=(u,v,w)$ is the velocity in the streamwise ($x$), wall-normal ($y$), and spanwise ($z$) directions, $p$ is the pressure, and $t$ is the time.
All quantities are nondimensionalized by the fluid density $\rho^*$, the channel half-width $\delta^*$, and the friction velocity $u_\tau^*$.
The superscript $(\cdot)^*$ denotes dimensional quantities, and $(\cdot)^+$ denotes quantities in wall units.
The friction Reynolds number is defined as $\Retau=u_\tau^*\delta^*/\nu^*$, where $\nu^*$ is the kinematic viscosity, and is set to $\Retau=110$.

\begin{figure}[tb]
 \centering
 \includegraphics[width=\linewidth]{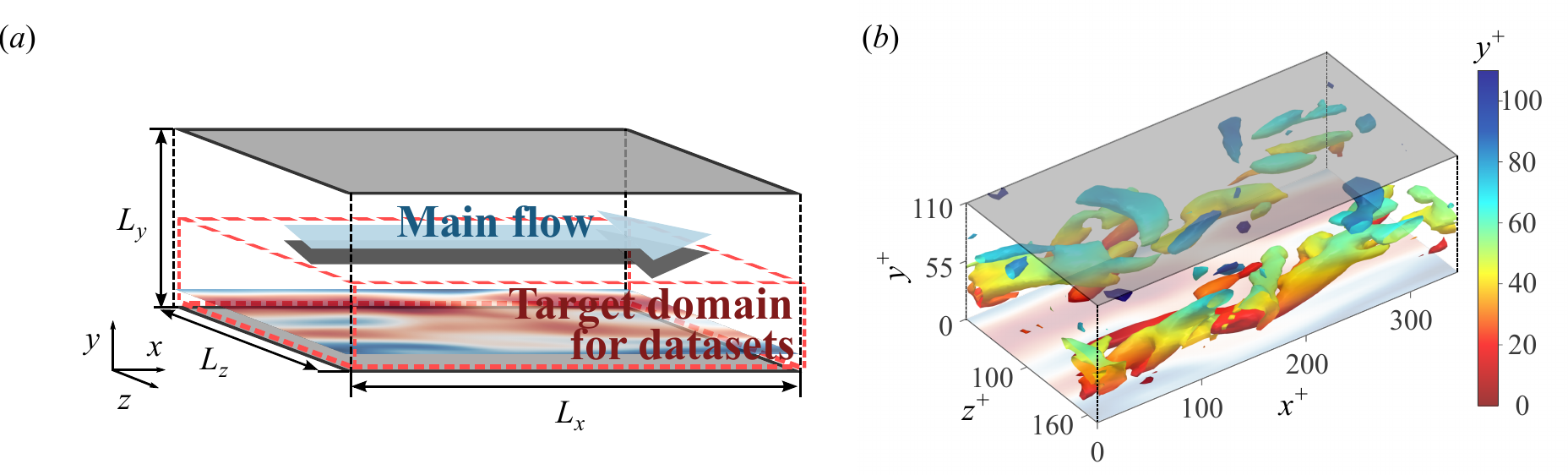}
 \caption{Minimal turbulent channel flow at $\mathrm{Re}_\tau=110$: (\textit{a}) computational domain, where the red dashed lines indicate the lower half of the channel used for the datasets; (\textit{b}) instantaneous vortical structures visualized by the isosurface of the second invariant of the velocity gradient tensor ($Q^+=0.01$), colored by $y^+$. The bottom plane shows the streamwise velocity at $y^+\simeq15.5$.}
 \label{fig:method_domain}
\end{figure}

Figure~\ref{fig:method_domain}(\textit{a}) shows the computational domain.
The domain size is $(L_x,L_y,L_z)=(\pi,2,\pi/2)$, or $(L_x^+,L_y^+,L_z^+)\simeq(346,220,173)$ in wall units, which is close to the minimal domain required for self-sustaining wall turbulence \citep{JimenezMoin1991}.
In such a domain, the low-order turbulence statistics in the near-wall region agree well with experiments, and the flow consists of a few streaks and quasi-streamwise vortices that sustain the near-wall turbulence \citep{JimenezMoin1991}.
The minimal channel retains the near-wall structures of channel turbulence with far fewer grid points than a large domain.
The minimal channel flow is thus a suitable target for examining the dimension of the latent space for wall turbulence.
The numbers of grid points are $(N_x,N_y,N_z)=(32,64,32)$, and the grid is uniform in all three directions.
The corresponding grid spacings in wall units are $(\Delta x^+,\Delta y^+,\Delta z^+)\simeq(10.8,3.44,5.40)$.
Periodic boundary conditions are imposed in the $x$ and $z$ directions, and the no-slip condition is imposed on the top and bottom walls.

Equations~\eqref{eq:method_continuity} and \eqref{eq:method_ns} are solved by a finite-difference DNS on a staggered grid.
For the spatial discretization, the energy-conservative fourth-order central difference scheme \citep{Morinishi1998} is used in the $x$ and $z$ directions, and the second-order central difference scheme is used in the $y$ direction.
For the time integration, the low-storage third-order Runge--Kutta method \citep{SpalartMoserRogers1991} is used for the convection terms and the viscous terms in the $x$ and $z$ directions, and the Crank--Nicolson method is used for the viscous terms in the $y$ direction.
The velocity and the pressure are coupled by a high-order scheme similar to the simplified marker and cell method \citep{DukowiczDvinsky1992}, and the Poisson equation for the pressure is solved with the fast Fourier transform in the $x$ and $z$ directions and a tridiagonal solver in the $y$ direction.
The time step is $\Delta t^+=0.0385$ in wall units.
The mean velocity and the root-mean-square (RMS) velocity fluctuations of the present DNS are compared with those of a spectral DNS in a large domain \citep{Iwamoto2002Database,Iwamoto2002DNS} in appendix~\ref{app:dns_correlation}.
Figure~\ref{fig:method_domain}(\textit{b}) shows instantaneous vortical structures visualized by the isosurface $Q^+=0.01$ of the second invariant $Q$ of the velocity gradient tensor.

\subsection{Datasets}
\label{subsec:method_data}

The datasets consist of the three-dimensional velocity fields of the minimal turbulent channel flow obtained by the DNS described above.
The sampling interval of the snapshots is $\Delta t^+=3.85$, which corresponds to 100 time steps of the DNS.
Because the velocity is computed on the staggered grid, it is interpolated to the cell centers by fourth-order Lagrange interpolation.
Considering the statistical symmetry about the channel center plane, we use the three velocity components in the lower half of the channel ($0\leq y\leq1$), indicated by the red dashed lines in figure~\ref{fig:method_domain}(\textit{a}).
The input of one snapshot is an array of size (32, 32, 32, 3), with $N_q=98\,304$ elements.

\begin{figure}[tb]
 \centering
 \includegraphics[width=\linewidth]{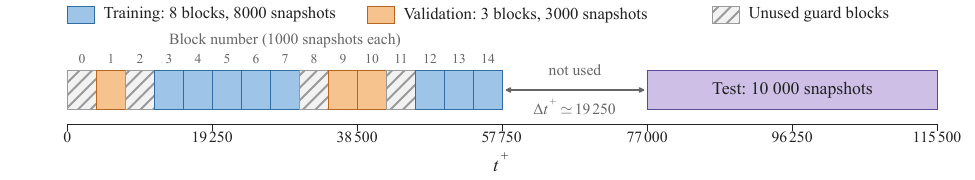}
 \caption{Temporal split of the DNS snapshots into the training, validation, and test data. The hatched guard blocks are not used.}
 \label{fig:method_split}
\end{figure}

Figure~\ref{fig:method_split} shows how the snapshots are divided into the training, validation, and test data.
The three datasets are separated to avoid correlation between snapshots that are close in time.
For the training and validation data, we use 15\,000 snapshots in the period $0<t^+\leq57\,750$ and divide them into 15 consecutive blocks of 1000 snapshots.
The blocks are numbered from 0 to 14 in time, and three randomly selected blocks (Nos.~1, 9, and 10; 3000 snapshots) are used as the validation data.
The four blocks adjacent to the validation blocks (Nos.~0, 2, 8, and 11) are left unused as guard blocks, and the remaining eight blocks (Nos.~3--7 and 12--14; 8000 snapshots) are used as the training data.
As a result, any training snapshot is at least one block, i.e., $\Delta t^+\simeq3850$, apart from any validation snapshot.
For the test data, we use 10\,000 snapshots in a separate period, $77\,004\leq t^+\leq115\,500$.
The interval between the end of the training and validation period and the start of the test period is $\Delta t^+\simeq19\,250$.
The autocorrelation of the bulk velocity, a slowly varying global quantity, first reaches zero at a time lag of $\Delta t^+\simeq3560$, and the autocorrelations of the near-wall velocity fluctuations decay faster (appendix~\ref{app:dns_correlation}).
Because both the interval between the training and validation data and that before the test data are longer than this time lag, the correlation between the three datasets can be regarded as small.
The trained models can thus be evaluated with test data that are almost uncorrelated with the data used for training.

The input and the target of the AE are the velocity fluctuations
\begin{equation}
 \bm q(x,y,z,t)=\bm u(x,y,z,t)-\bm U(y),
 \label{eq:method_fluctuation}
\end{equation}
where $\bm U(y)$ is the velocity profile averaged over time and the $x$ and $z$ directions of the training data.
The same $\bm U(y)$ is subtracted from the validation and test data.
No normalization is applied to the velocity components; its effect on the reconstruction of $v'$ and $w'$ is discussed in section~\ref{subsec:results_physics}.

\subsection{Three-dimensional CNN-AE}
\label{subsec:method_cnn}

\begin{figure}[tb]
 \centering
 \includegraphics[width=\linewidth]{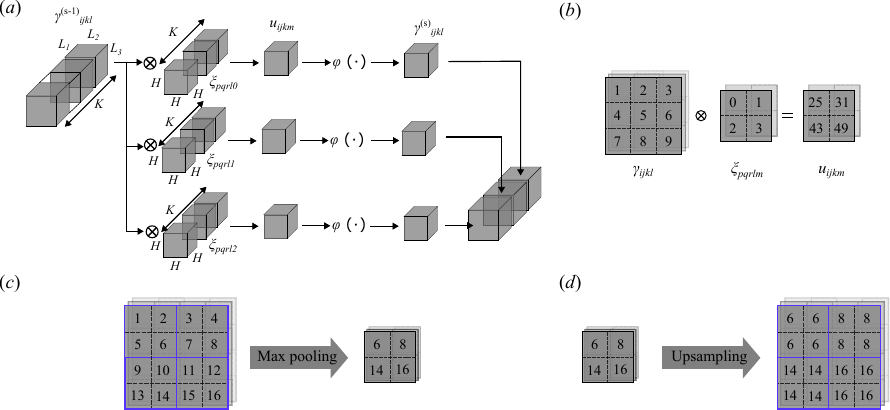}
 \caption{Operations used in the CNN-AE: (\textit{a}) three-dimensional convolutional layer; (\textit{b}) local filter operation; (\textit{c}) max pooling; (\textit{d}) upsampling. Panels (\textit{b})--(\textit{d}) show two-dimensional examples.}
 \label{fig:method_cnn_operations}
\end{figure}

An AE is trained to minimize the difference between the reconstruction $\hat{\bm q}$ and the input $\bm q$.
The encoder and decoder of a CNN-AE are built from convolution, max pooling, and upsampling operations, which are illustrated in figure~\ref{fig:method_cnn_operations}.
The convolution that gives the output $\gamma^{(s)}$ of the $s$-th layer is written as
\begin{equation}
 \gamma^{(s)}_{ijkm}=\varphi\!\left(
 b^{(s)}_m+
 \sum_{l=0}^{L-1}\sum_{p=0}^{H-1}\sum_{q=0}^{H-1}\sum_{r=0}^{H-1}
 \xi^{(s)}_{pqrlm}\,
 \gamma^{(s-1)}_{i+p-G,\,j+q-G,\,k+r-G,\,l}
 \right),
 \label{eq:method_conv}
\end{equation}
where $(i,j,k)$ are the indices of the grid points of the output, $l$ and $m$ are the indices of the input and output channels, $L$ is the number of input channels, $H$ is the kernel width, and $G=\mathrm{floor}(H/2)$.
The kernel $\xi^{(s)}$ and the bias $b^{(s)}$ are trainable, and $\varphi$ is the activation function.
Near the boundaries, we use periodic padding in the $x$ and $z$ directions, which supplies the values from the opposite side \citep{Guastoni2021Periodic}, and zero padding in the $y$ direction.
Owing to the periodic padding, the convolution retains the neighboring relation imposed by the periodic boundary conditions of the DNS.
The effect of the periodic padding is shown in appendix~\ref{app:padding}.

\begin{figure}[tb]
 \centering
 \includegraphics[width=\linewidth]{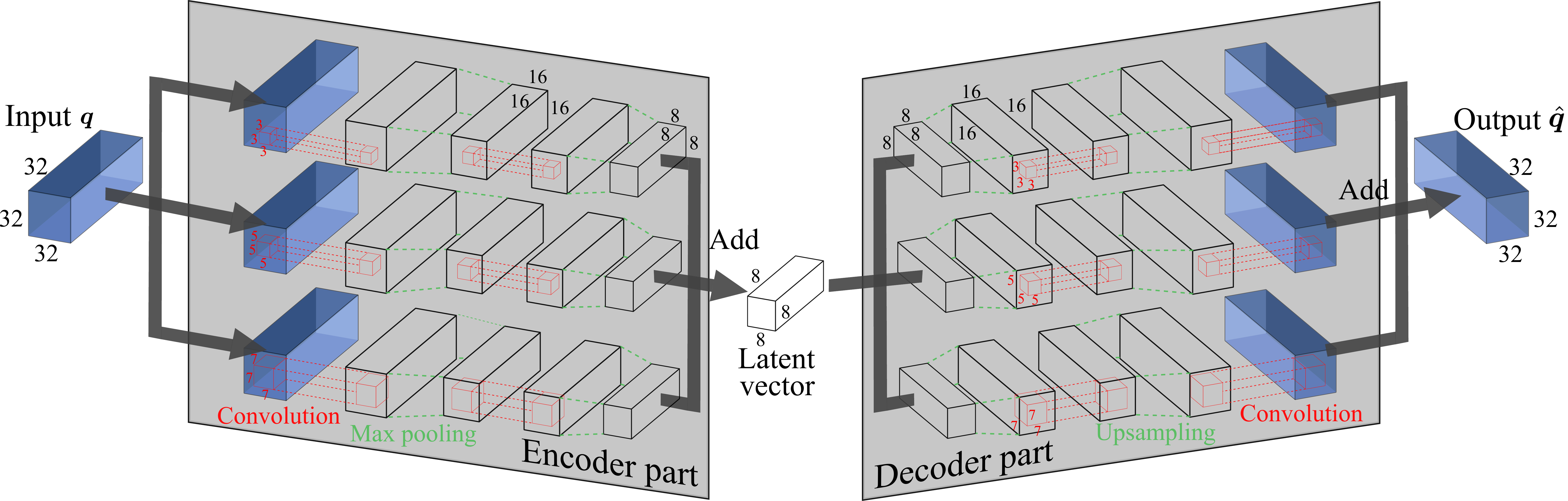}
 \caption{Architecture of the three-dimensional multi-scale CNN-AE.}
 \label{fig:method_cnn_backbone}
\end{figure}

Following the previous study on the minimal channel flow \citep{Nakamura2021Channel}, we use the three-dimensional multi-scale CNN-AE shown in figure~\ref{fig:method_cnn_backbone}, which is based on the multi-scale CNN used for the super-resolution reconstruction of turbulent flows \citep{Fukami2019SuperResolution}.
Because the size of the region covered by one convolution is determined by the kernel width, a small kernel captures fine structures, whereas a large kernel captures structures extending over a wide region.
As shown in figure~\ref{fig:method_cnn_backbone}, the multi-scale CNN-AE has three parallel branches with kernel widths of $H=3$, 5, and 7, and each branch processes the same input separately.
Because a turbulent field contains structures of different sizes, this model is suitable for capturing the various structures in the field.
Table~\ref{tab:method_cnn} lists the layers of each branch.
In each branch of the encoder, convolution and max pooling are repeated to reduce the number of grid points to $1/4$ in each direction, and the outputs of the three branches are added element-wise.
In this way, the features extracted from regions of different sizes are combined into one feature map on the same $8\times8\times8$ grid.
This feature map is flattened to give the output of the encoder,
\begin{equation}
 \bm h=\mathcal E_{\mathrm{CNN}}(\bm q)\in\mathbb R^{F},\qquad F=8^3C,
 \label{eq:method_flatten}
\end{equation}
where $C$ is the number of channels on the coarsest grid, and $F$ is the dimension of $\bm h$, hereafter referred to as the width.
We mainly use $F=1536$ and 3072, i.e., $C=3$ and 6.
As listed in table~\ref{tab:method_cnn}, the decoder restores the number of grid points in the reverse order of the encoder, adds the outputs of the three branches in the same way, and outputs the three velocity components with a final convolution.
Hereafter, the decoder that maps $\bm h$ to $\hat{\bm q}$ is denoted by $\mathcal D_{\mathrm{CNN}}$.
The model that passes $\bm h$ directly to $\mathcal D_{\mathrm{CNN}}$,
\begin{equation}
 \hat{\bm q}=\mathcal D_{\mathrm{CNN}}\bigl(\mathcal E_{\mathrm{CNN}}(\bm q)\bigr),
 \label{eq:method_baseline}
\end{equation}
is hereafter referred to as the baseline CNN-AE.

\begin{table}[tb]
 \centering
 \caption{Structure of each branch of the multi-scale CNN-AE (kernel width $H\in\{3,5,7\}$).}
 \label{tab:method_cnn}
 \small
 \begin{tabular}{@{}lll@{\hspace{1.2em}}lll@{}}
  \toprule
  \multicolumn{3}{c}{Encoder $\mathcal E_\mathrm{CNN}$} & \multicolumn{3}{c}{Decoder $\mathcal D_\mathrm{CNN}$} \\
  \cmidrule(r{1.2em}){1-3}\cmidrule(l{0pt}){4-6}
  Layer & Output shape & Activation & Layer & Output shape & Activation \\
  \cmidrule[\lightrulewidth](r{1.2em}){1-3}\cmidrule[\lightrulewidth](l{0pt}){4-6}
  Input $\bm q$ & (32, 32, 32, 3) & --- & Input $\bm h$ & ($F$) & --- \\
  Conv. & (32, 32, 32, 16) & ReLU & Reshape & (8, 8, 8, $C$) & --- \\
  Conv. & (32, 32, 32, 8) & ReLU & Conv. & (8, 8, 8, $C$) & ReLU \\
  Max pooling & (16, 16, 16, 8) & --- & Upsampling & (16, 16, 16, $C$) & --- \\
  Conv. & (16, 16, 16, 16) & ReLU & Conv. & (16, 16, 16, 8) & ReLU \\
  Conv. & (16, 16, 16, 8) & ReLU & Conv. & (16, 16, 16, 16) & ReLU \\
  Max pooling & (8, 8, 8, 8) & --- & Upsampling & (32, 32, 32, 16) & --- \\
  Conv. & (8, 8, 8, $C$) & tanh & Conv. & (32, 32, 32, 8) & ReLU \\
  Add (3 branches) & (8, 8, 8, $C$) & --- & Conv. & (32, 32, 32, 16) & ReLU \\
  Flatten & ($F$) & --- & Add (3 branches) & (32, 32, 32, 16) & --- \\
  & & & Conv.\ ($3\times3\times3$) & (32, 32, 32, 3) & linear \\
  \bottomrule
 \end{tabular}
\end{table}

\subsection{IRMAE-WD}
\label{subsec:method_irmae}

\begin{figure}[tb]
 \centering
 \includegraphics[width=\linewidth]{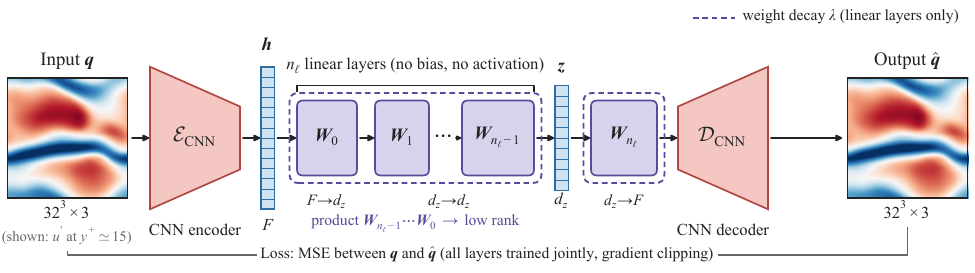}
 \caption{Schematic of IRMAE-WD.}
 \label{fig:method_irmae}
\end{figure}

In the present study, we use IRMAE-WD \citep{Zeng2024IRMAEWD}, which inserts linear layers without activation functions or biases between $\bm h$ and $\mathcal D_{\mathrm{CNN}}$ of the baseline CNN-AE.
With $\nl$ linear layers on the encoder side, as shown in figure~\ref{fig:method_irmae}, the latent variables $\bm z$ and the reconstruction $\hat{\bm q}$ are given by
\begin{equation}
 \bm z=\bm W_{\nl-1}\cdots\bm W_1\bm W_0\,\bm h\in\mathbb R^{\dz},
 \qquad
 \hat{\bm q}=\mathcal D_{\mathrm{CNN}}(\bm W_{\nl}\bm z),
 \label{eq:method_irmae}
\end{equation}
where $\bm h$ is the output of the CNN encoder ($F$-dimensional), and $\bm z$ is the latent variable obtained by mapping $\bm h$ through the linear layers.
The dimension $\dz$ of $\bm z$ is hereafter referred to as the latent width, and $\dz\leq F$ in the present study.
The matrix $\bm W_0$ is $\dz\times F$, the matrices $\bm W_1,\ldots,\bm W_{\nl-1}$ are $\dz\times\dz$, and $\bm W_{\nl}$ is the $F\times\dz$ matrix that maps the latent variables back to $F$ dimensions.

The AE is trained to find the weights that minimize the squared reconstruction error.
We denote by $\bm w$ all the kernels and biases of the CNN and the matrices of the linear layers, and by $\hat{\bm q}=\mathcal F(\bm q;\bm w)$ the whole model that maps the input $\bm q$ to $\hat{\bm q}$ (equation~\eqref{eq:method_baseline} for the baseline CNN-AE and equation~\eqref{eq:method_irmae} for IRMAE-WD).
The training then solves
\begin{equation}
 \bm w^\ast=\mathop{\rm arg\,min}_{\bm w}\frac{1}{N\,N_q}\sum_{n=1}^{N}\bigl\|\bm q_n-\mathcal F(\bm q_n;\bm w)\bigr\|_2^2,
 \label{eq:method_argmin}
\end{equation}
where $N$ is the number of training snapshots, and $N_q$ is the number of elements in one snapshot.
In practice, the training data are divided into mini-batches of 128 snapshots, and the weights are updated with the gradient of the mean squared error (MSE) for the mini-batch $\mathcal B_s$ used in the $s$-th update,
\begin{equation}
 \mathcal L_s=\frac{1}{|\mathcal B_s|\,N_q}\sum_{n\in\mathcal B_s}\|\bm q_n-\hat{\bm q}_n\|_2^2,
 \label{eq:method_loss}
\end{equation}
which serves as the loss function.
We update the weights with AdamW \citep{LoshchilovHutter2019}.
AdamW is a variant of Adam \citep{KingmaBa2015}, which adjusts the step size of each weight from moving averages of the past gradients and their magnitudes, and it applies the weight decay to all weights at the same rate, separately from the adaptive update.
Adam itself is widely used, but when weight decay is added to the loss function as $L_2$ regularization, the gradient of the decay term is normalized together with the loss gradient by the square root of the second-moment estimate.
The decay is then weaker for weights with larger gradients and nonuniform among the weights.
\citet{Zeng2024IRMAEWD} also reported that AdamW is important for IRMAE-WD and that the results become sensitive to the parameters when Adam is used with $L_2$ regularization.
In addition, to avoid abrupt updates by large gradients, the gradient $\bm g_s$ of $\mathcal L_s$ is clipped by its norm over all parameters \citep{Pascanu2013},
\begin{equation}
 \widetilde{\bm g}_s=\frac{c}{\max(c,\|\bm g_s\|_2)}\,\bm g_s,\qquad c=1.
 \label{eq:method_clip}
\end{equation}
In other words, the gradient is scaled down to the norm $c$, without changing its direction, only when its norm exceeds the threshold $c$.
The weight $\bm W$ of each linear layer is updated as
\begin{equation}
 \bm W^{(s+1)}=(1-\alpha\lambda)\,\bm W^{(s)}-\alpha\,\mathcal A_s(\widetilde{\bm g}_s)_{\bm W},
 \label{eq:method_adamw}
\end{equation}
where $\alpha$ is the learning rate, $\lambda$ is the weight decay coefficient, and $\mathcal A_s(\cdot)_{\bm W}$ is the part of the adaptive update direction of Adam that corresponds to $\bm W$.
The factor $(1-\alpha\lambda)$ in the first term on the right-hand side is the decay that shrinks the weights of the linear layers at each iteration.
The decay is applied only to $\bm W_0,\ldots,\bm W_{\nl}$ and not to the kernels and biases of the CNN.
The reasons for this choice and for the gradient clipping are given in appendix~\ref{app:stabilization}, and the effect of the weight decay coefficient $\lambda$ is shown in appendix~\ref{app:weight_decay}.

\subsection{Training procedures}
\label{subsec:method_training}

\begin{figure}[tb]
 \centering
 \includegraphics[width=\linewidth]{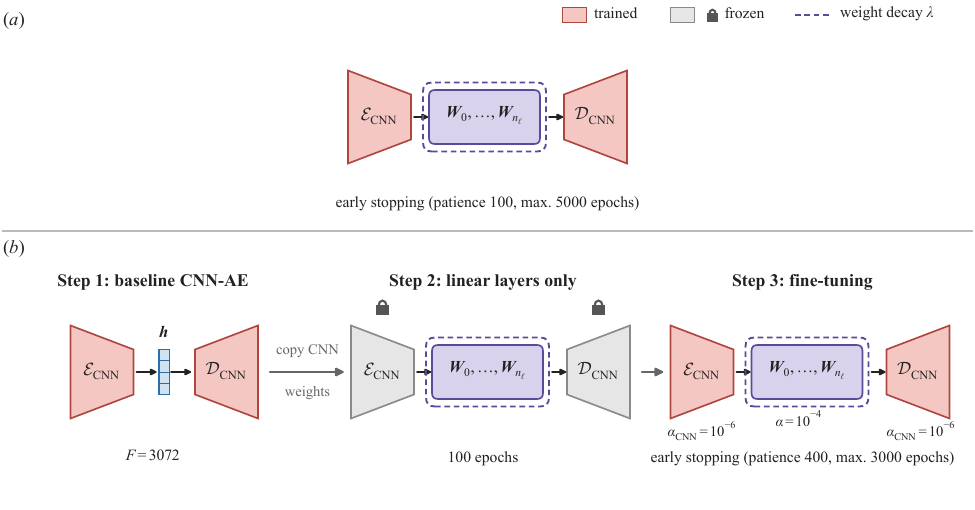}
 \caption{Training procedures: (\textit{a}) training from random initialization; (\textit{b}) transfer learning from the trained baseline CNN-AE in three steps. Dashed outlines indicate the layers to which weight decay is applied.}
 \label{fig:method_training}
\end{figure}

In this subsection, we describe the two procedures used to train the models, which differ in the initial values of the weights (figure~\ref{fig:method_training} and table~\ref{tab:method_training}).
In the first procedure, all weights are initialized randomly; this procedure is hereafter referred to as training from random initialization.
In the second, the CNN is initialized with the weights of a trained baseline CNN-AE; this procedure is hereafter referred to as transfer learning.

In training from random initialization (figure~\ref{fig:method_training}(\textit{a})), models with $F=1536$ are trained, and for IRMAE-WD the latent width is $\dz=768$--$1536$ and the number of linear layers is $\nl=2$, 3, 4, and 6.
All weights are initialized with the Glorot uniform distribution \citep{Glorot2010}, and each of the 12 configurations listed in table~\ref{tab:method_training} is trained three times with different random seeds, which change the initial weights and the order of the mini-batches.
The learning rate is $\alpha=10^{-3}$, and the weight decay coefficient is $\lambda=10^{-3}$.
Training is stopped when the validation loss has not improved for 100 epochs (at most 5000 epochs), and the weights with the lowest validation loss are used.

In transfer learning (figure~\ref{fig:method_training}(\textit{b})), models with $F=3072$ are trained, and for IRMAE-WD the latent width is $\dz=F=3072$ and the number of linear layers is $\nl=2$, 4, 6, 8, and 10.
The CNN of IRMAE-WD is initialized with the weights of a trained baseline CNN-AE, which has no linear layers.
This procedure is used because the IRMAE-WD models with $F\geq3072$ trained from random initialization lost accuracy, and those with more linear layers collapsed to a solution that outputs only the mean field (appendix~\ref{app:stabilization}).
Transfer learning, which reuses the weights of a trained model as the initial values for training another model, is widely used \citep{Yosinski2014}.
In fluid dynamics, it has been used, for example, to apply a model trained with data at a low Reynolds number to data at a higher Reynolds number \citep{Inubushi2020Transfer,Guastoni2021Periodic}.
Initial weights obtained by training have also been used to make the training feasible when it fails from random initial weights: \citet{HintonSalakhutdinov2006} showed that deep autoencoders can be trained successfully when the initial weights are close to a good solution.
The transfer learning consists of three steps.
In step 1, the baseline CNN-AE with $F=3072$ is trained under the same conditions as in training from random initialization.
In step 2, linear layers initialized with the Glorot uniform distribution are inserted between the trained $\mathcal E_{\mathrm{CNN}}$ and $\mathcal D_{\mathrm{CNN}}$, and only the linear layers are trained for 100 epochs with the learning rate $\alpha=10^{-4}$, while the CNN weights are frozen, i.e., not updated.
The CNN is frozen in this step to prevent the trained CNN weights from being destroyed by large gradients passing through the randomly initialized linear layers.
In machine learning, it has also been shown that when new layers are added to a pretrained model, training only these layers first and then fine-tuning the whole model better preserves the pretrained features \citep{Kumar2022LPFT}.
In step 3, starting from the weights with the lowest validation loss in step 2, all layers are fine-tuned with a newly initialized optimizer.
The learning rate is $\alpha=10^{-4}$ for the linear layers and $\alpha_{\mathrm{CNN}}=10^{-6}$ for the CNN, without learning-rate decay.
The small learning rate for the CNN keeps its weights from moving far from the trained values; the learning rate was reduced for the same reason in transfer learning for wall turbulence \citep{Guastoni2021Periodic}.
Training is stopped when the validation loss has not improved for 400 epochs (at most 3000 epochs), and the weights with the lowest validation loss are used for evaluation.
For $\nl=4$ and 6, however, training was continued to 3000 epochs without early stopping, to examine how the drop position changes as training proceeds.
The weight decay coefficient is $\lambda=10^{-3}$, and in both steps 2 and 3 the weight decay is applied only to the linear layers.
The computing environment, the numerical precision, and the training time are given in appendix~\ref{app:environment}.

\begin{table}[tb]
 \centering
 \caption{Training conditions.}
 \label{tab:method_training}
 \small
 \begin{tabular}{lll}
  \toprule
  & Random initialization & Transfer learning \\
  \midrule
  Width $F$ of $\bm h$ & 1536 & 3072 \\
  IRMAE-WD $(d_z, n_\ell)$ & $d_z\in\{768,1024,1280\}$, $n_\ell\in\{2,3\}$; & $d_z=3072$, $n_\ell\in\{2,4,6,8,10\}$ \\
  & $d_z=768$, $n_\ell=4$; $d_z=1024$, $n_\ell\in\{4,6\}$; & \\
  & $d_z=1536$, $n_\ell\in\{2,3,4\}$ & \\
  Seeds & 3 & 1 \\
  Initialization & Glorot uniform & CNN: baseline CNN-AE; linear: Glorot \\
  Learning rate & $\alpha=10^{-3}$ & linear $\alpha=10^{-4}$, CNN $\alpha_\mathrm{CNN}=10^{-6}$ \\
  $\lambda$ (IRMAE-WD) & $10^{-3}$ & $10^{-3}$ \\
  Epochs & early stopping (patience 100, max.\ 5000) & 100 (step 2) + early stopping (patience 400, \\
  & & max.\ 3000; step 3) \\
  Weights evaluated & lowest validation loss & lowest validation loss in step 3 \\
  \bottomrule
 \end{tabular}
\end{table}

\subsection{Singular value decomposition and truncation of the latent variables}
\label{subsec:method_svd}

Although the latent variable $\bm z$ of IRMAE-WD has $\dz$ components (figure~\ref{fig:method_irmae}), the latent variables of all training snapshots may lie only in a subspace of lower dimension than $\dz$.
In this case, the number of coordinates that the model actually uses to represent the velocity fields is not $\dz$ but the dimension of this subspace, which we refer to as the effective dimension of the representation, $d_{\mathrm{eff}}$.
In this subsection, we use the singular value decomposition (SVD) of the latent variables to determine $d_{\mathrm{eff}}$, which is compared in section~\ref{sec:results} with the intrinsic degrees of freedom of the flow.

\begin{figure}[tb]
 \centering
 \includegraphics[width=\linewidth]{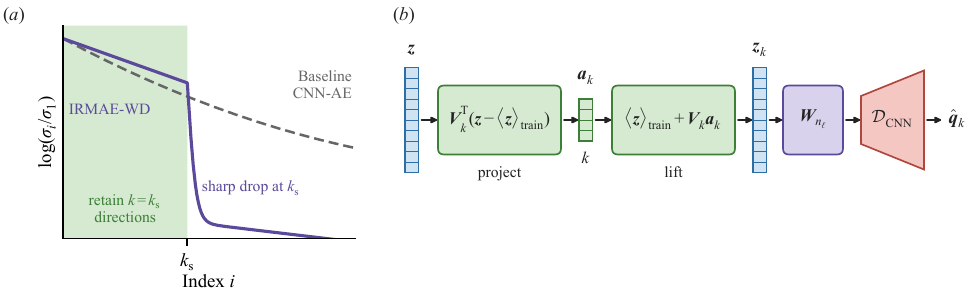}
 \caption{Latent singular value decomposition and truncation: (\textit{a}) singular values of the centered latent variables of the training data (schematic); (\textit{b}) reconstruction truncated to $k$ directions with all trained weights fixed.}
 \label{fig:method_svd}
\end{figure}

The SVD decomposes the matrix of the centered latent variables of the training data as
\begin{equation}
 \bm Z_c=\begin{bmatrix}
  (\bm z_1-\langle\bm z\rangle_{\mathrm{train}})^{\mathsf T}\\ \vdots\\
  (\bm z_{N}-\langle\bm z\rangle_{\mathrm{train}})^{\mathsf T}
 \end{bmatrix}=\bm U\bm\Sigma\bm V^{\mathsf T},
 \label{eq:method_svd}
\end{equation}
where $\bm z_n$ is the latent variable obtained by feeding the $n$-th of the $N=8000$ training snapshots ($n=1,\ldots,N$) to the trained model, and $\langle\bm z\rangle_{\mathrm{train}}$ is their mean, where $\langle\cdot\rangle_{\mathrm{train}}$ denotes the average over the training snapshots.
The matrices $\bm U$ and $\bm V$ have orthonormal columns, and $\bm\Sigma=\mathrm{diag}(\sigma_1,\sigma_2,\ldots,\sigma_{\dz})$ is the diagonal matrix of the singular values.
The columns of $\bm V$ are mutually orthogonal directions in the latent space, and the $i$-th singular value $\sigma_i$ ($\sigma_1\geq\sigma_2\geq\cdots\geq\sigma_{\dz}$) measures the spread of the latent variables along the $i$-th direction.
A direction with a larger $\sigma_i$ carries more of the variation among the snapshots.
If the latent variables lie only in a subspace spanned by fewer directions than $\dz$, the singular values beyond that number are nearly zero, and the singular values drop sharply around the index corresponding to that number (figure~\ref{fig:method_svd}(\textit{a})).
This index is the drop position.
\citet{Zeng2024IRMAEWD} estimated the number of coordinates needed to represent the data from this drop position.
For the minimal channel flow, we examine in section~\ref{subsec:results_spectrum} whether the drop position gives $d_{\mathrm{eff}}$.

The directions beyond the drop position carry little of the variance of the latent variables and are expected to contribute little to the reconstruction.
To verify this, we use truncation, in which only the first $k$ directions in descending order of $\sigma_i$ are kept and the velocity field is reconstructed with the same trained decoder,
\begin{equation}
 \bm a_k=\bm V_k^{\mathsf T}(\bm z-\langle\bm z\rangle_{\mathrm{train}}),\qquad
 \bm z_k=\langle\bm z\rangle_{\mathrm{train}}+\bm V_k\bm a_k,\qquad
 \hat{\bm q}_k=\mathcal D_{\mathrm{CNN}}(\bm W_{\nl}\bm z_k),
 \label{eq:method_truncation}
\end{equation}
as illustrated in figure~\ref{fig:method_svd}(\textit{b}).
Here, $k$ is the number of retained directions, i.e., the number of coordinates used for the reconstruction, $\bm V_k$ consists of the first $k$ columns of $\bm V$, and $\bm a_k\in\mathbb R^k$ are the $k$ coordinates.
The case using all directions of the latent variables without truncation ($k=\dz$, or $k=F$ for the baseline CNN-AE) is referred to as the full width.
If the drop position marks $d_{\mathrm{eff}}$, the reconstruction rate (equation~\eqref{eq:method_eta}) should increase with $k$ and hardly change for $k$ larger than the drop position.
This is confirmed for the actual models in section~\ref{subsec:results_warmstart}.
To state the number of coordinates required for a given accuracy, we also define $k_{90}$ and $k_{95}$ as the smallest $k$ for which the reconstruction rate on the validation data reaches 90\% and 95\%, respectively.
Because a representation truncated to $k$ directions uses exactly $k$ coordinates, $k_{90}$ and $k_{95}$ can be regarded as the effective dimensions of the representation needed to reach these accuracies.
To keep the test data independent of these choices, the mean and the directions of the latent space are computed from the training data only, $k_{90}$ and $k_{95}$ are selected with the validation data only, and the test data are used only to report the performance at the selected $k$.

The drop position is determined in the following two ways, depending on the training procedure.
For the models trained from random initialization, $\log_{10}(\sigma_i/\sigma_1)$ is fitted by two straight lines against the index $i$, and the boundary index that minimizes the sum of the squared residuals, with at least five points on each line, is taken as the drop position $\ks$.
The sharpness of the drop is measured by the number of decades by which the singular values decrease over the 100 directions centered at $\ks$,
\begin{equation}
 \Delta_{100}=\log_{10}\frac{\sigma_{\ks-50}}{\sigma_{\ks+50}}.
 \label{eq:method_drop}
\end{equation}
For the models obtained by transfer learning, the spectra fall off steeply at their end, and for models without a clear drop the boundary of the two-line fit is placed at the first bend of the decrease, near the end of the spectrum, or on a flat part.
For these models, we instead define the drop position $\kd$ as the index at which the drop is sharpest.
That is, $\kd$ is the index $i$ in the range $512\leq i\leq2500$ that maximizes the decrease in equation~\eqref{eq:method_drop} with $\ks$ replaced by $i$, and $\Delta_{100}$ at $\kd$ gives the sharpness of the drop.
The range is set to exclude the first bend of the spectra at small indices and the fall-off near their end.
The two definitions are compared in appendix~\ref{app:knee}, together with further details on the drop position.

\subsection{Methods compared with IRMAE-WD}
\label{subsec:method_comparison}

In the present study, the baseline CNN-AE and POD are used as references for IRMAE-WD.
To compare the truncation results and the required numbers of coordinates with those of IRMAE-WD, we describe in this subsection how each of the two methods reconstructs the velocity field from $k$ coordinates.
Both methods are evaluated with the same data and metrics as IRMAE-WD (section~\ref{subsec:method_metrics}).

The first is the baseline CNN-AE.
We regard $\bm h$ in equation~\eqref{eq:method_flatten} itself as the latent variable, apply the same SVD and truncation as in section~\ref{subsec:method_svd}, and pass the result to the same $\mathcal D_{\mathrm{CNN}}$.
In equation~\eqref{eq:method_truncation}, $\bm z$ is replaced by $\bm h$, and $\bm W_{\nl}$ by the identity map.
This comparison shows the number of directions of the CNN-AE latent variables needed for the reconstruction without linear layers.

The second is POD.
From the $N=8000$ training snapshots, we compute the mean field $\langle\bm q\rangle_{\mathrm{train}}$ and the matrix of the fluctuations,
\begin{equation}
 \bm Q=\bigl[\bm q_1-\langle\bm q\rangle_{\mathrm{train}},\;\bm q_2-\langle\bm q\rangle_{\mathrm{train}},\;\ldots,\;\bm q_N-\langle\bm q\rangle_{\mathrm{train}}\bigr]\in\mathbb R^{N_q\times N}.
 \label{eq:method_snapshot}
\end{equation}
The POD modes are the eigenvectors of the covariance matrix $\bm Q\bm Q^{\mathsf T}$.
Because $\bm Q\bm Q^{\mathsf T}$ is an $N_q\times N_q$ matrix with $N_q=98\,304$, solving its eigenvalue problem directly is costly.
We use the method of snapshots \citep{Sirovich1987}, which instead solves the eigenvalue problem of the much smaller $N\times N$ matrix $\bm Q^{\mathsf T}\bm Q$,
\begin{equation}
 \bm Q^{\mathsf T}\bm Q\,\bm\psi_i=\lambda^{\mathrm{POD}}_i\bm\psi_i,
 \label{eq:method_snapshot_eig}
\end{equation}
and obtains the orthonormal POD modes as
\begin{equation}
 \bm\phi_i=\frac{\bm Q\bm\psi_i}{\bigl(\lambda^{\mathrm{POD}}_i\bigr)^{1/2}}\qquad(i=1,2,\ldots),
 \label{eq:method_snapshot_mode}
\end{equation}
where $\lambda^{\mathrm{POD}}_1\geq\lambda^{\mathrm{POD}}_2\geq\cdots$ and each $\bm\psi_i$ has unit length.
The eigenvalue $\lambda^{\mathrm{POD}}_i$ is the fluctuation energy of the training data captured by the mode $\bm\phi_i$.
A snapshot $\bm q$ of the validation or test data is then projected onto the first $k$ modes, and the field is reconstructed from the $k$ coefficients as
\begin{equation}
 \hat{\bm q}^{\mathrm{POD}}_k=\langle\bm q\rangle_{\mathrm{train}}+\sum_{i=1}^{k}\bigl[\bm\phi_i^{\mathsf T}(\bm q-\langle\bm q\rangle_{\mathrm{train}})\bigr]\bm\phi_i,
 \label{eq:method_pod}
\end{equation}
where $\bm\phi_i^{\mathsf T}(\bm q-\langle\bm q\rangle_{\mathrm{train}})$ is the inner product over all grid points and components.
Because the modes and the mean field are computed from the training data only, the validation and test data are not used to construct the basis, and $k_{90}$ and $k_{95}$ of POD are determined by the same rule as for IRMAE-WD.
Since POD is the optimal linear representation of the training data, this comparison shows whether the nonlinear mapping of an AE reduces the number of coordinates needed for a given accuracy.

\subsection{Evaluation metrics}
\label{subsec:method_metrics}

We evaluate the reconstructions with the overall reconstruction rate, the relative error of each component, the ratio of the number of coordinates to that of POD, the RMS velocity fluctuations, and the scale-wise reconstruction rate.

The overall reconstruction rate $\eta$ is defined as
\begin{equation}
 \eta=1-\frac{\sum_n\|\bm q_n-\hat{\bm q}_n\|_2^2}
              {\sum_n\|\bm q_n-\langle\bm q\rangle_{\mathrm{train}}\|_2^2},
 \label{eq:method_eta}
\end{equation}
where the sums are taken over the snapshots $n$ of the evaluated data (the validation data when $k_{90}$ and $k_{95}$ are selected, and the test data otherwise), and $\langle\bm q\rangle_{\mathrm{train}}$ is the mean field of the training data.
Although $\bm q$ is the fluctuation from $\bm U(y)$, the mean $\langle\bm q\rangle_{\mathrm{train}}$ computed from a finite number of snapshots is not exactly zero.
We subtract $\langle\bm q\rangle_{\mathrm{train}}$ in the denominator as well, consistent with the POD reconstruction (equation~\eqref{eq:method_pod}), which is based on $\langle\bm q\rangle_{\mathrm{train}}$.
A reconstruction that returns only $\langle\bm q\rangle_{\mathrm{train}}$ (POD with $k=0$) then gives $\eta=0$, and an exact reconstruction gives $\eta=1$.
The rate $\eta$ represents the fraction of the variation about the mean field of the training data that is reconstructed.
However, $\eta$ is dominated by the errors of the components with large fluctuations, and the accuracy of each component should be evaluated separately.
For this purpose, we define the relative error $\epsilon_c$ of each component as
\begin{equation}
 \epsilon_c=\frac{\sum_n\|q_{c,n}-\hat q_{c,n}\|_2^2}{\sum_n\|q_{c,n}-\langle q_c\rangle_{x,z,t}\|_2^2},\qquad c=u,v,w,
 \label{eq:method_component}
\end{equation}
where $\langle\cdot\rangle_{x,z,t}$ denotes the average over time and the $x$ and $z$ directions of the evaluated data, and the denominator measures the magnitude of the fluctuations of component $c$.

To compare the efficiency of a representation with that of POD, we use the ratio $k/k^\mathrm{POD}$ of the numbers of coordinates required for the same reconstruction rate.
Here, $k^\mathrm{POD}$ is the smallest number of POD modes with which POD reaches, on the test data, the reconstruction rate $\eta(k)$ of the AE truncated to $k$ directions,
\begin{equation}
 k^\mathrm{POD}=\min\bigl\{k'\bigm|\eta^\mathrm{POD}(k')\geq\eta(k)\bigr\},
 \label{eq:method_kpod}
\end{equation}
where $\eta^\mathrm{POD}(k')$ is the reconstruction rate of POD with $k'$ modes.
With this definition, a ratio $k/k^\mathrm{POD}$ smaller than one means that the AE reaches the same reconstruction rate as POD with fewer coordinates.

As turbulence statistics, we compute the wall-normal profiles of the RMS velocity fluctuations,
\begin{equation}
 q'_{c,\mathrm{rms}}(y)=\bigl\langle q_c'^2\bigr\rangle_{x,z,t}^{1/2},\qquad q'_c=q_c-\langle q_c\rangle_{x,z,t},
 \label{eq:method_rms}
\end{equation}
for the reconstructed fields and the DNS, where the averages are taken over all test snapshots.
These are denoted by $u^{+\prime}_\mathrm{rms}$, $v^{+\prime}_\mathrm{rms}$, and $w^{+\prime}_\mathrm{rms}$ in wall units for $c=u$, $v$, and $w$.

The scale-wise reconstruction rate $R_c$, defined below, is used to examine the relationship between the number of retained coordinates and the spatial scales that can be reproduced.
For the fluctuation $q_c'$ of component $c$ in equation~\eqref{eq:method_rms}, the Fourier coefficient $\widetilde q_c$ in the $x$ direction is
\begin{equation}
 \widetilde q_c(\kappa_x;y,z,t)=\frac{1}{N_x}\sum_{i=0}^{N_x-1}q_c'(x_i,y,z,t)\,e^{-\mathrm i\kappa_xx_i},\qquad
 \kappa_x=\frac{2\pi m_x}{L_x}\quad(m_x=0,1,\ldots,N_x/2),
 \label{eq:method_fourier}
\end{equation}
and the one-dimensional energy spectrum $E_c(\kappa_x)$ is defined as
\begin{equation}
 E_c(\kappa_x)=a_{m_x}\sum_{j}\bigl\langle|\widetilde q_c(\kappa_x;y_j,z,t)|^2\bigr\rangle_{z,t},
 \label{eq:method_spectrum}
\end{equation}
where $c=u,v,w$ denotes the velocity component, $\langle\cdot\rangle_{z,t}$ denotes the average over the $z$ direction and the test snapshots, and the sum is taken over the grid points $y_j$ in the lower half of the channel.
The coefficient $a_{m_x}$ accounts for the contribution of the negative wavenumbers and is 1 for $m_x=0$ and $N_x/2$ and 2 otherwise.
Similarly, the spectrum computed from the error field $\hat q_c'-q_c'$ is denoted by $E^{\mathrm{err}}_c(\kappa_x)$, and the scale-wise reconstruction rate is defined as
\begin{equation}
 R_c(\lambda_x)=1-\frac{E^{\mathrm{err}}_c(\kappa_x)}{E_c(\kappa_x)},\qquad \lambda_x=\frac{2\pi}{\kappa_x}.
 \label{eq:method_scale_ratio}
\end{equation}
The same definitions are used in the $z$ direction.
Here, $\lambda_x$ and $\lambda_z$ are the streamwise and spanwise wavelengths, i.e., the sizes of the structures in these directions.
The wavelengths that can be represented in the present domain are $\lambda_x^+=346/m_x$ and $\lambda_z^+=173/m_z$ ($m_x,m_z=1,\ldots,16$), and the component with $\kappa_x=0$ ($m_x=0$) is uniform in the streamwise direction.
In section~\ref{subsec:results_scale}, $R_c$ is used to judge whether the structures of each wavelength are reproduced with $k$ coordinates.
Because $R_c$ is computed from the spectrum of the error field, it counts not only amplitude errors but also phase errors.
From $R_c$, we also define the smallest wavelength $\lambda_{\min}(k)$ that can be reproduced with $k$ coordinates: for a reconstruction truncated to $k$ directions, $\lambda_{\min}(k)$ is the smallest wavelength such that $R_c(\lambda)\geq\theta$ for all wavelengths $\lambda\geq\lambda_{\min}$, with the threshold $\theta=0.5$.
Because $\theta=0.3$ and 0.7 lead to almost the same comparison between IRMAE-WD and POD, $\theta=0.5$ is used here as an example.
In addition, to examine the scales at which the errors remain, we use the share of the error energy $E^{\mathrm{err}}_{\mathcal B}$ in a streamwise wavelength band $\mathcal B$ in the total error and its ratio to the total fluctuation energy $E_{\mathrm{tot}}$, where
\begin{equation}
 E^{\mathrm{err}}_{\mathcal B}=\sum_{c}\sum_{\kappa_x\in\mathcal B}E^{\mathrm{err}}_c(\kappa_x),\qquad
 E_{\mathrm{tot}}=\sum_{c}\sum_{\kappa_x}E_c(\kappa_x).
 \label{eq:method_band}
\end{equation}
The spanwise wavelength bands are defined in the same way.
\section{Results and discussion}
\label{sec:results}

\subsection{Drop position and its meaning}
\label{subsec:results_spectrum}
\label{subsec:results_tradeoff}

In this subsection, we examine IRMAE-WD trained from random initialization: whether a drop appears in the latent spectrum, what determines its position, and what the drop position means.

\begin{figure}[tb]
 \centering
 \includegraphics[width=\linewidth]{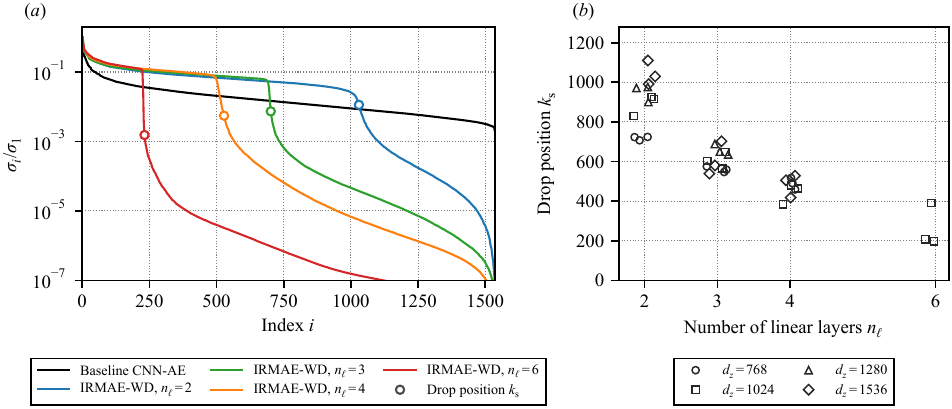}
 \caption{Latent singular value spectra of IRMAE-WD trained from random initialization: (\textit{a}) normalized singular values for $d_z=1536$ and $n_\ell=2$, 3, 4, and 6 (seed 1) and for the baseline CNN-AE ($F=1536$); (\textit{b}) drop position $k_\mathrm{s}$ of all 36 models against the number of linear layers $n_\ell$.}
 \label{fig:results_spectra}
\end{figure}

Figure~\ref{fig:results_spectra}(\textit{a}) shows the normalized singular values $\sigma_i/\sigma_1$ of the latent variables of the training data for the models with $\dz=1536$ and $\nl=2$, 3, 4, and 6 (seed 1) and for the baseline CNN-AE ($F=1536$).
For all IRMAE-WD models, the singular values decrease gradually and then drop sharply at a certain index.
In contrast, the singular values of the baseline CNN-AE, which has no linear layers, decrease gradually to the end without a drop.
For all 36 models, $\Delta_{100}$, defined in equation~\eqref{eq:method_drop}, ranges from 1.3 to 5.1, i.e., the singular values decrease by more than one decade across the 100 directions around the drop position. The drop positions and reconstruction rates of all 69 models, including the configurations trained with seed 1 only (e.g., $\nl=8$ and $\dz=192$--$512$), are summarized in appendix~\ref{app:random_table}.
Thus, a clear drop appears in the latent spectra of IRMAE-WD for the three-dimensional turbulent velocity field as well.

Figure~\ref{fig:results_spectra}(\textit{b}) shows the drop position $\ks$ against the number of linear layers $\nl$ for all 12 configurations and three seeds.
The drop position decreases systematically as linear layers are added, from about 700--1100 for $\nl=2$ to about 200--400 for $\nl=6$.
This trend is consistent with the property that a product of matrices is more strongly biased toward low-rank solutions as the number of layers increases \citep{Arora2019DeepMF}, and the same trend has been reported for two-dimensional Rayleigh--B\'enard convection \citep{Vinograd2025RBC}.
For small $\nl$, the drop position also depends on the latent width: for $\nl=2$, $\ks$ increases with the latent width and reaches 93\% of the latent width for $\dz=768$, whereas it depends only weakly on the latent width for $\nl\geq3$.
That is, for $\nl=2$, the drop position is limited by the latent width.
This is probably because the implicit regularization by a product of only two linear layers is weak, and the model uses most of the available directions.
In addition, the drop position changes much less with the weight decay coefficient than with the number of linear layers (appendix~\ref{app:weight_decay}).
The drop position is thus determined not by the flow alone but by the network configuration, i.e., mainly by the number of linear layers and, for $\nl=2$, also by the latent width.

\begin{figure}[tb]
 \centering
 \includegraphics[width=\linewidth]{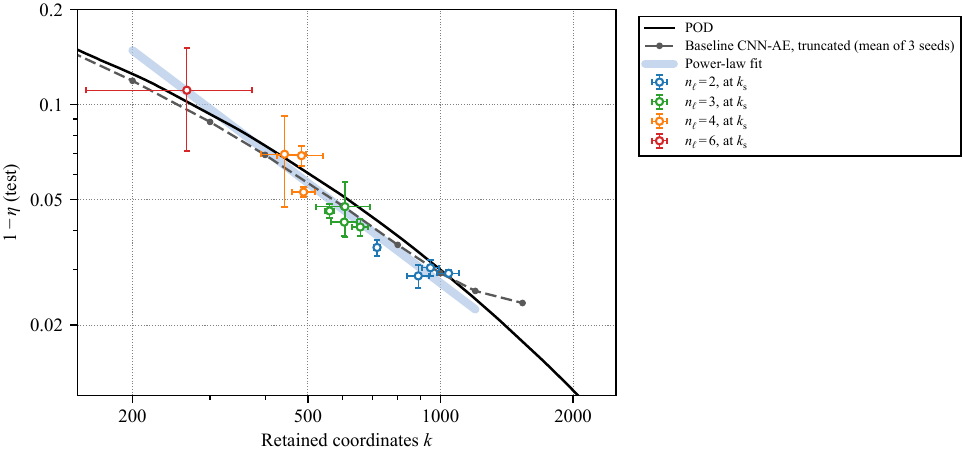}
 \caption{Models trained from random initialization ($F=1536$): test error against the number of retained coordinates $k$ when each IRMAE-WD model is truncated at its drop position (mean and standard deviation over three seeds), compared with the baseline CNN-AE truncated to $k$ directions (mean over three seeds) and POD.}
 \label{fig:results_tradeoff}
\end{figure}

We next examine the reconstruction accuracy when the latent variables are truncated at the drop position.
When each model is truncated to $k=\ks$ with the directions obtained from the latent variables of the training data, the reconstruction rate on the test data decreases by 0.002 percentage points on average over the 36 models and by at most 0.01 percentage points.
This shows that the directions beyond the drop position contribute little to the reconstruction of the trained model, and the drop position accurately indicates $d_{\mathrm{eff}}$.
To compare this accuracy among different numbers of linear layers, figure~\ref{fig:results_tradeoff} shows the reconstruction error $1-\eta$ on the test data against the number of retained coordinates $k$.
The points for IRMAE-WD truncated at $\ks$ (open circles) move toward fewer coordinates and larger errors as linear layers are added.
Thus, adding linear layers not only reduces the drop position but also lowers the reconstruction accuracy.
This is presumably because, with more linear layers, the variations of the latent variables passed to the decoder are confined to fewer directions, and the variations that cannot be represented in these directions are lost in the reconstruction.
A similar degradation with too many linear layers has also been reported for Rayleigh--B\'enard convection \citep{Vinograd2025RBC}.

Regardless of the number of linear layers, these points lie close to a single straight line on logarithmic axes, approximated by $1-\eta\simeq39.5\,k^{-1.05}$.
For $k\lesssim1000$, this line nearly coincides with the curve (dashed line) obtained by applying the same SVD and truncation to the latent variable $\bm h$ of the baseline CNN-AE with the same width ($F=1536$), which has no linear layers.
Hence, for this flow, the drop position provides no more information than the truncation of the baseline CNN-AE.
Furthermore, the reconstruction error decreases continuously with the number of coordinates, and no threshold appears beyond which the error becomes abruptly small.
Note that the dashed curve of the baseline CNN-AE flattens for $k\gtrsim1000$ because it approaches the full-width accuracy of the model itself (97.65\% on average over the three seeds), not because of a threshold of the flow.
As described in section~\ref{sec:introduction}, when the solutions lie on a low-dimensional manifold, as for the Kuramoto--Sivashinsky equation, the reconstruction error hardly decreases once the latent width exceeds the manifold dimension, and the drop position coincides with this dimension \citep{LinotGraham2020,Zeng2024IRMAEWD}.
In contrast, for the present velocity field, no such threshold appears in the relationship between the error and the number of coordinates, even though the latent spectrum exhibits a drop, and it is unclear which directions are unnecessary for the reconstruction.
This suggests that the drop position is not uniquely determined by the flow, but is set by the strength of the implicit regularization, which depends on the number of linear layers.
In other words, the drop position of a single trained model cannot be regarded as the intrinsic degrees of freedom of the flow.
We conjecture that no direction is clearly unnecessary because the minimal channel flow contains fluctuations over a continuous and wide range of scales, from large-scale structures to small eddies, and every direction contributes a little to the reconstruction.
This point is examined in detail in section~\ref{subsec:results_scale}.

Next, we compare IRMAE-WD truncated at the drop position with POD by the ratio $k/k^\mathrm{POD}$, where $k^\mathrm{POD}$ is the number of POD modes required for the same test error.
For most configurations with $\nl=2$ and 3, the ratio is 0.83--0.93, i.e., IRMAE-WD requires up to 17\% fewer coordinates than POD for the same error.
These ratios are similar to or slightly smaller than those of the baseline CNN-AE compared at the same accuracy (0.90--0.98).
In contrast, for some configurations with more linear layers ($\nl=4$ and 6), the ratio is 1.03--1.12, i.e., more coordinates than POD are required.
Even for $\nl=2$, the ratio exceeds one at around $\eta\simeq97\%$.
In short, for the configurations with fewer linear layers ($\nl=2$ and 3), the representations truncated at the drop position reach the same error as POD with fewer coordinates for $\eta\lesssim97\%$, but their advantage over POD decreases as the required $\eta$ increases.

\subsection{Latent spectra and number of coordinates of the models obtained by transfer learning}
\label{subsec:results_warmstart}

The IRMAE-WD models in the previous subsection, with $F=1536$, reach reconstruction rates of at most 97.35\%.
To examine higher reconstruction rates, we present in this subsection the latent spectra and the required number of coordinates for models with a larger width, $F=3072$.
IRMAE-WD models with $F\geq3072$ trained from random initialization, however, suffered from degraded accuracy or collapse (appendix~\ref{app:stabilization}).
Instead, we present the results obtained by transfer learning from a trained baseline CNN-AE with $F=3072$.

\begin{figure}[tb]
 \centering
 \includegraphics[width=\linewidth]{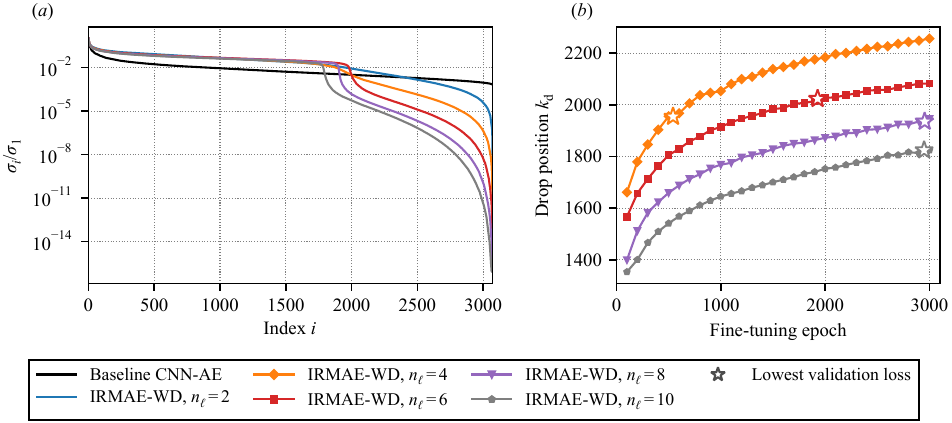}
 \caption{Latent spectra of the models obtained by transfer learning: (\textit{a}) normalized singular values (weights with the lowest validation loss); (\textit{b}) drop position $k_\mathrm{d}$ against the fine-tuning epoch.}
 \label{fig:results_transfer}
\end{figure}

We first examine whether, as for the models trained from random initialization, a drop appears in the latent spectra of the models obtained by transfer learning and whether its position indicates $d_{\mathrm{eff}}$.
Figure~\ref{fig:results_transfer}(\textit{a}) shows the normalized singular values of the latent variables of the training data for each model.
The spectrum of the baseline CNN-AE decreases gradually to the end without a clear drop.
Because the baseline CNN-AE has no linear layers that bias it toward low rank, its latent variables are expected to use all directions of the width.
For IRMAE-WD with $\nl=2$, the drop is also unclear, and the largest decrease lies at the edge of the search range.
This is probably because, as for the models trained from random initialization (section~\ref{subsec:results_spectrum}), the implicit regularization by a product of only two linear layers is weak, and the transfer learning, in which the CNN is almost fixed, inherits the spectrum of the features of the baseline CNN-AE, which has no drop.
For IRMAE-WD with $\nl\geq4$, on the other hand, a drop appears, and it becomes sharper as linear layers are added from $\nl=4$ to 8.
Thus, in transfer learning as well, a drop appears in the latent spectrum with four or more linear layers.
Note, however, that the best weights of the models are taken at different epochs and that the drop position moves during training, as shown below; the comparison of the best weights reflects both the number of linear layers and the training progress.
In particular, for $\nl=8$ and 10, the lowest validation loss was reached near the end of training (epochs 2954 and 2949), i.e., the training had not fully converged at 3000 epochs.
When the models with $\nl=4$--10 are truncated at the drop position, the reconstruction rate decreases by at most 0.004 percentage points, and the directions beyond the drop position are hardly used for the reconstruction.
That is, as for the models trained from random initialization, the drop position of the models obtained by transfer learning indicates $d_{\mathrm{eff}}$.
Next, figure~\ref{fig:results_transfer}(\textit{b}) shows the drop position $\kd$ against the epoch in step 3 of the transfer learning (fine-tuning of all layers).
For every $\nl$, the drop position moves almost monotonically toward more coordinates as training proceeds, and the drop becomes sharper at the same time ($\Delta_{100}$ increases from 0.24--0.85 at epoch 100 to 1.08--1.98 at epoch 3000).
At every epoch, the drop position decreases in the order $\nl=4$, 6, 8, and 10.
Thus, although the drop position changes as training proceeds, the trend that adding linear layers strengthens the implicit regularization and moves the drop position to fewer coordinates holds in transfer learning as well.

\begin{figure}[tb]
 \centering
 \includegraphics[width=\linewidth]{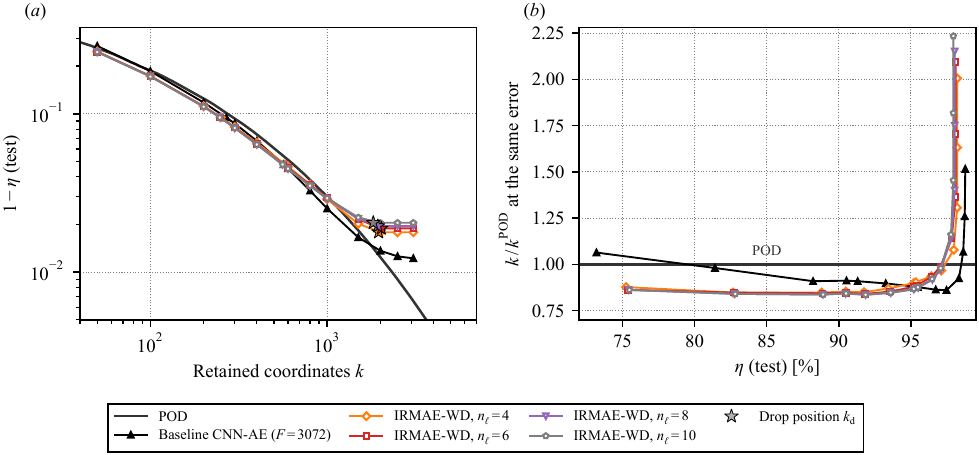}
 \caption{Models obtained by transfer learning ($F=3072$): (\textit{a}) test error after truncation to $k$ directions, where the stars mark the drop position $k_\mathrm{d}$; (\textit{b}) ratio of $k$ to $k^\mathrm{POD}$ at the same test error. IRMAE-WD with $n_\ell=2$, which has no drop, is listed in table~\ref{tab:results_transfer} only.}
 \label{fig:results_transfer_curves}
\end{figure}

\begin{table}[tb]
 \centering
 \caption{Representations obtained by transfer learning ($F=d_z=3072$, seed 1) and POD. ``---'' for $k_\mathrm{d}$: no drop, i.e., the largest decrease over 100 directions lies at the edge of the search range. $^\dagger$POD with $k=3072$ modes. Epoch: epoch with the lowest validation loss in step 3.}
 \label{tab:results_transfer}
 \small
 \begin{tabular}{lrrrrrrr}
  \toprule
  Representation & Epoch & $k_{90}$ & $k_{95}$ & $\eta$ (full) [\%] & $k_\mathrm{d}$ & $\Delta_{100}$ & $k^\mathrm{POD}$ \\
  \midrule
  POD & --- & 296 & 649 & 99.34$^\dagger$ & --- & --- & --- \\
  Baseline CNN-AE & --- & 269 & 567 & 98.78 & --- & 0.07 & 2027 \\
  IRMAE-WD, $n_\ell=2$ & 392 & 251 & 562 & 98.32 & --- & 0.17 & 1607 \\
  IRMAE-WD, $n_\ell=4$ & 542 & 250 & 580 & 98.21 & 1954 & 0.46 & 1532 \\
  IRMAE-WD, $n_\ell=6$ & 1928 & 248 & 567 & 98.11 & 2023 & 1.50 & 1467 \\
  IRMAE-WD, $n_\ell=8$ & 2954 & 246 & 558 & 98.05 & 1935 & 1.97 & 1428 \\
  IRMAE-WD, $n_\ell=10$ & 2949 & 247 & 559 & 97.95 & 1824 & 1.86 & 1375 \\
  \bottomrule
 \end{tabular}
\end{table}

We next examine the number of coordinates needed to reach a given accuracy when each model is truncated, and compare it with those of POD and the baseline CNN-AE.
First, figure~\ref{fig:results_transfer_curves}(\textit{a}) shows the test error when each model is truncated from $k=50$ to the full width.
For IRMAE-WD models with a drop, the error hardly changes for $k$ larger than the drop position $\kd$ (stars).
Thus, truncating the model at the drop position gives almost the same accuracy as the full width, and retaining more coordinates hardly improves the accuracy.
Table~\ref{tab:results_transfer} summarizes, for the baseline CNN-AE, IRMAE-WD ($\nl=2$--10), and POD, the smallest numbers of coordinates $k_{90}$ and $k_{95}$ at which the reconstruction rate on the validation data reaches 90\% and 95\%, the full-width $\eta$, and the drop position and its sharpness.
Here, $k^\mathrm{POD}$ is the number of POD modes required for the same test error as the full-width model.
As listed in table~\ref{tab:results_transfer}, a 90\% reconstruction requires 246--251 coordinates for IRMAE-WD, compared with 296 modes for POD.
Figure~\ref{fig:results_transfer_curves}(\textit{b}) shows the ratio $k/k^\mathrm{POD}$ at the same test error.
The ratio of IRMAE-WD is smaller than one for $\eta\lesssim97\%$, regardless of the number of linear layers.
Compared with the baseline CNN-AE, the ratios of IRMAE-WD with $\nl=4$--10 are all smaller for $\eta\lesssim93\%$; at $\eta=90\%$, for example, IRMAE-WD reaches the same error with 7--8\% fewer coordinates than the baseline CNN-AE.
These results show that IRMAE-WD obtained by transfer learning reaches the same error as POD with fewer coordinates for $\eta\lesssim97\%$.
In particular, for $\eta\lesssim93\%$, it requires 12--16\% fewer coordinates than POD and reduces the dimension more efficiently than the baseline CNN-AE.
For $k\lesssim1000$, the curves for $\nl=4$--10 nearly coincide, and $k_{90}$ and $k_{95}$ for $\nl=2$, which has no drop, are almost the same as well (table~\ref{tab:results_transfer}).
That is, the effective dimension required for a given accuracy is almost the same regardless of the number of linear layers.
Hence, regardless of the presence or position of the drop, the effective dimension for the desired reconstruction accuracy can be determined by truncating a single IRMAE-WD model.

As $\eta$ increases, on the other hand, the ratio approaches one.
Above $\eta\simeq97\%$, POD reaches the same error with fewer coordinates than IRMAE-WD, and above $\eta\simeq98.5\%$, also than the baseline CNN-AE.
In addition, the number of coordinates at the drop position exceeds the number of POD modes $k^\mathrm{POD}$ required for the same error.
The full-width reconstruction rates of both AEs are also lower than that of POD with the same 3072 modes (table~\ref{tab:results_transfer}).
This advantage of POD probably arises because the accuracy of the baseline CNN-AE, from which IRMAE-WD is initialized, saturates below that of POD.
The scales at which the errors remain at this saturation are examined in section~\ref{subsec:results_scale}.
In short, for $\eta\gtrsim97\%$ ($k\gtrsim1000$), POD represents the velocity field with fewer coordinates than IRMAE-WD.
However, even with all 8000 modes available from the training data, POD reaches a reconstruction rate of 99.87\% on the test data.
In other words, the modes beyond the first 3072 raise the reconstruction rate by only 0.53 percentage points.
This is because the remaining fluctuation energy is spread thinly over a large number of modes.
Thus, for the minimal channel flow, an accuracy as high as $\eta\gtrsim99\%$ requires many coordinates even with POD.

\subsection{Statistics of the reconstructed turbulent fields}
\label{subsec:results_physics}

In this subsection, we examine whether the truncated reconstructions are physically reasonable as turbulent fields by using the statistics of the velocity fluctuations.

\begin{figure}[tb]
 \centering
 \includegraphics[width=\linewidth]{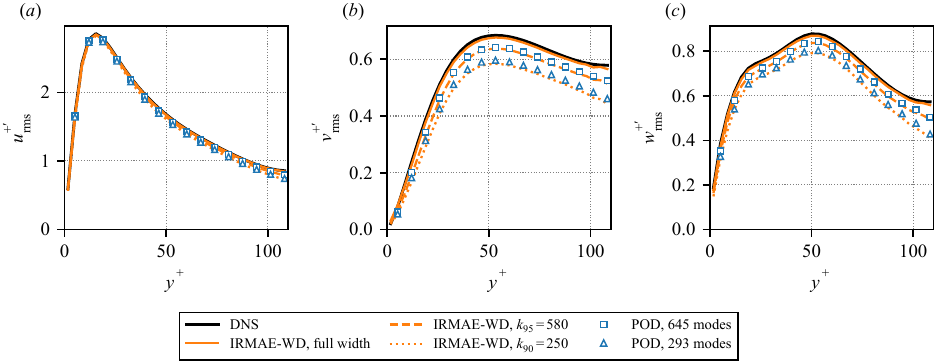}
 \caption{Wall-normal profiles of the root-mean-square velocity fluctuations on the test data: (\textit{a}) $u'$, (\textit{b}) $v'$, and (\textit{c}) $w'$, for IRMAE-WD with $n_\ell=4$ obtained by transfer learning and for POD with the numbers of modes (645 and 293) that give the same test error as IRMAE-WD truncated to $k_{95}$ and $k_{90}$.}
 \label{fig:results_rms}
\end{figure}

Figure~\ref{fig:results_rms} compares the wall-normal profiles of the RMS velocity fluctuations reconstructed by IRMAE-WD with $\nl=4$ obtained by transfer learning at the full width, $k_{95}$, and $k_{90}$ with those of the DNS.
The symbols show the profiles of POD with the numbers of modes (645 and 293) that give the same test error as IRMAE-WD truncated to $k_{95}$ and $k_{90}$.
The RMS of $u'$ agrees well with the DNS, including the peak around $y^+\simeq15$, even when the model is truncated to $k_{90}$.
The profiles of all three components are almost the same as those of POD at the same reconstruction rate.
The profiles of the turbulent kinetic energy and the Reynolds shear stress are also almost the same as those of POD at the same reconstruction rate (appendix~\ref{app:turb_stats}).
These results show that the truncated reconstructions with IRMAE-WD reproduce statistics such as the RMS velocity fluctuations, the turbulent kinetic energy, and the Reynolds shear stress comparably to POD at the same reconstruction rate.

On the other hand, the RMS of $v'$ and $w'$ decreases at all wall-normal locations as the number of coordinates is reduced, and at $k_{90}$ the RMS of $v'$ is clearly smaller than that of the DNS.
In terms of the relative error of each component, $\epsilon_v=0.29$ and $\epsilon_w=0.21$ at $k_{90}$, compared with $\epsilon_u=0.059$; the relative errors of $v'$ and $w'$ are three to five times that of $u'$.
Nevertheless, the overall reconstruction rate reaches $\eta=90\%$.
This is because the components are not normalized by their amplitudes in the present study, and $u'$ accounts for about 78\% of the sum of the squared fluctuations in the test data, whereas $v'$ and $w'$ account for only about 8\% and 13\%, respectively.
The same tendency is found for the baseline CNN-AE and for IRMAE-WD with other numbers of linear layers, and the relative errors of POD at the same reconstruction rate are almost the same as those of IRMAE-WD.
Since an AE is trained to minimize the squared reconstruction error and POD gives the linear subspace that minimizes the same squared error, both are expected to give priority to the reconstruction of $u'$, which carries most of the fluctuation energy.
The fact that the overall reconstruction rate remains high despite the large errors of $v'$ and $w'$ is thus not specific to AEs, but is likely common to representations evaluated with a squared error that is not normalized by the amplitude of each component.
Accordingly, when the number of coordinates needed to represent the flow is discussed, the errors of the individual components should be checked in addition to the overall reconstruction rate.

\subsection{Reproducible spatial scales and the required number of coordinates}
\label{subsec:results_scale}

So far, we have shown that the reconstruction error decreases continuously with the number of coordinates and that the required number of coordinates depends on the required accuracy.
For reduced-order modeling, however, the size of the structures that can be reproduced also matters.
The RMS examined in section~\ref{subsec:results_physics} does not show this size, because it agrees with the DNS for $u'$ even at $k_{90}$ and is averaged over time and the $x$ and $z$ directions.
In this subsection, we pair the required number of coordinates with the spatial scales that can be reproduced, and we also examine at which scales the errors of IRMAE-WD remain at high accuracy (section~\ref{subsec:results_warmstart}) and why no direction is clearly unnecessary (section~\ref{subsec:results_spectrum}).

\begin{figure}[!b]
 \centering
 \includegraphics[width=\linewidth]{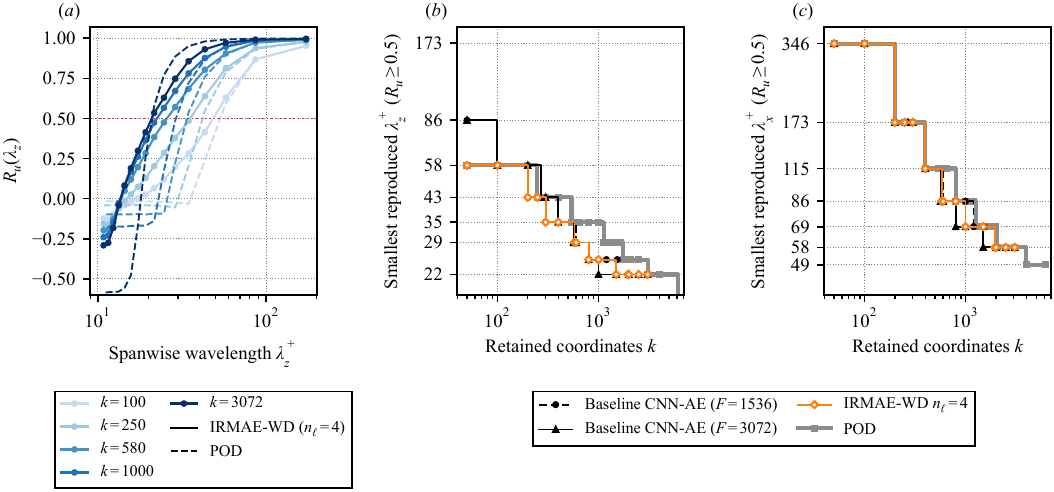}
 \caption{Scale-wise reconstruction of $u'$ on the test data: (\textit{a}) scale-wise reconstruction rate $R_u(\lambda_z)$ of IRMAE-WD with $n_\ell=4$ (solid) and POD (dashed) for several $k$; (\textit{b}) smallest spanwise and (\textit{c}) streamwise wavelengths reproduced with $R_u\geq0.5$ against $k$ for IRMAE-WD with $n_\ell=4$, the baseline CNN-AEs, and POD.}
 \label{fig:results_scale}
\end{figure}

Figure~\ref{fig:results_scale}(\textit{a}) shows the spanwise scale-wise reconstruction rate $R_u(\lambda_z)$ of $u'$ for IRMAE-WD with $\nl=4$ obtained by transfer learning and for POD.
For both representations, $R_u$ is close to one at long wavelengths and decreases as the wavelength becomes shorter; that is, large structures are reproduced accurately, whereas small structures are difficult to reproduce.
As the number of coordinates increases, $R_u$ at each wavelength increases, except at the shortest wavelengths.
That is, the more coordinates are retained, the smaller the structures that can be reproduced.

\begin{figure}[!t]
 \centering
 \includegraphics[width=\linewidth]{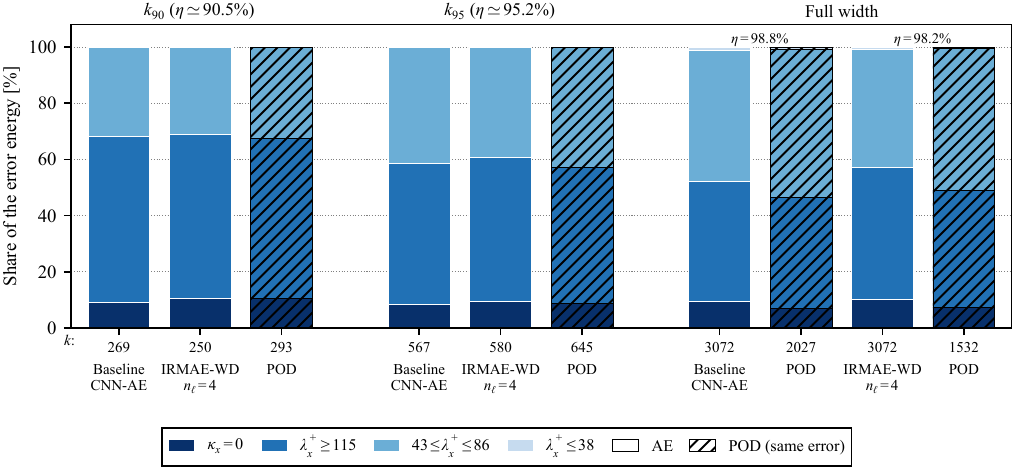}
 \caption{Share of the error energy in streamwise wavelength bands at $k_{90}$, $k_{95}$, and the full width. POD (hatched) has the same test error as IRMAE-WD in the $k_{90}$ and $k_{95}$ groups and as each AE at the full width. The number of retained coordinates $k$ is shown below each bar.}
 \label{fig:results_bands}
\end{figure}

Figures~\ref{fig:results_scale}(\textit{b}) and (\textit{c}) show the smallest wavelength $\lambda_{\min}(k)$ reproduced with $R_u\geq\theta=0.5$ (red dotted line in figure~\ref{fig:results_scale}(\textit{a})) against the number of coordinates $k$.
In the spanwise direction, IRMAE-WD with $\nl=4$ reproduces the structures with $\lambda_z^+\geq29$ with 580 coordinates, whereas POD requires more than 1000 modes.
This difference is larger than that for the same test error (figure~\ref{fig:results_transfer_curves}(\textit{b})).
In the streamwise direction as well, IRMAE-WD requires no more coordinates than POD to reproduce structures of the same wavelength.
Although the results are not shown, IRMAE-WD with $\nl=6$--10 requires almost the same numbers of coordinates as $\nl=4$ to reproduce the structures with $\lambda_z^+\geq25$ and $\lambda_x^+\geq69$.
However, IRMAE-WD cannot reproduce $\lambda_x^+\leq49$ even at the full width.
POD with the same 3072 modes does not reproduce these structures either; it reproduces $\lambda_x^+\simeq49$ with 4000 modes.
This suggests that 3072 coordinates are not sufficient to reproduce these structures, even with IRMAE-WD.
Compared with the baseline CNN-AE with $F=3072$, from which IRMAE-WD is initialized, IRMAE-WD reproduces the same scales with the same or fewer coordinates for $k\lesssim600$ ($\lambda_z^+\gtrsim29$).
For smaller scales, on the other hand, the baseline CNN-AE sometimes requires fewer coordinates.
This boundary is close to the range of $k$ ($k\simeq400$--580, $\eta\simeq93$--95\%) in which the ratio of IRMAE-WD with $\nl=4$ and that of the baseline CNN-AE cross in figure~\ref{fig:results_transfer_curves}(\textit{b}).
These results show that the required number of coordinates is uniquely determined once the scale to be reproduced and the threshold are specified.
Within the scales that IRMAE-WD can reproduce, this number is smaller for IRMAE-WD than for POD, and the difference increases as smaller scales are targeted.

We next examine why POD reaches the same error with fewer coordinates at high accuracy.
Figure~\ref{fig:results_bands} shows the share of the error energy in each streamwise wavelength band for the truncations at $k_{90}$, $k_{95}$, and the full width.
At $k_{90}$ ($\eta\simeq90.5\%$), the shares of IRMAE-WD and POD are almost the same.
As $\eta$ increases, however, the share of the error in the streamwise-uniform component ($\kappa_x=0$) and in the large-scale structures with $\lambda_x^+\geq115$ becomes larger for IRMAE-WD than for POD.
At the full width, this share reaches 57\% for IRMAE-WD with $\nl=4$, compared with 49\% for POD.
The baseline CNN-AE shows the same tendency, with 52\% compared with 47\% for POD at the full width.
In contrast, the share of the smallest wavelength band ($\lambda_x^+\leq38$) is about 1\% or less for both the AEs and POD, i.e., the errors remaining in small-scale structures contribute little to the total error.
In addition, as shown in figures~\ref{fig:results_scale}(\textit{b}) and (\textit{c}), IRMAE-WD reproduces smaller structures than POD with the same number of coordinates.
These results indicate that POD is more accurate at high $\eta$ not because the AEs fail to reproduce small-scale structures, but because they reconstruct the large-scale structures, which carry most of the energy, less precisely than POD.

\begin{figure}[!t]
 \centering
 \includegraphics[width=\linewidth]{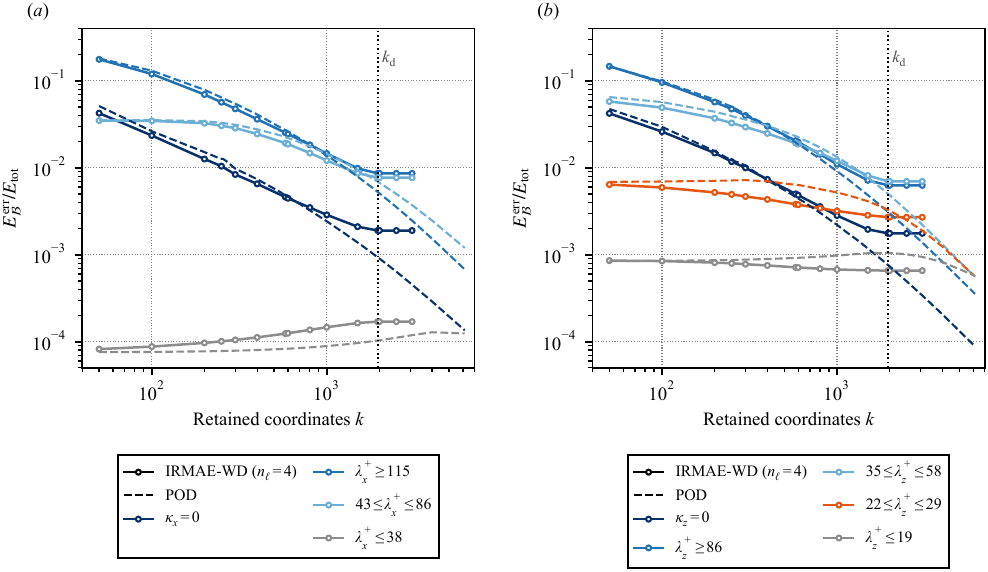}
 \caption{Error energy in wavelength bands against the number of retained coordinates $k$ for IRMAE-WD with $n_\ell=4$ obtained by transfer learning (solid) and POD (dashed): (\textit{a}) streamwise bands; (\textit{b}) spanwise bands.}
 \label{fig:results_band_error}
\end{figure}

We now consider, from the viewpoint of scales, why no direction is clearly unnecessary, as conjectured in section~\ref{subsec:results_spectrum}.
Figure~\ref{fig:results_band_error} shows the ratio $E^{\mathrm{err}}_{\mathcal B}/E_{\mathrm{tot}}$ of the error energy in each streamwise and spanwise wavelength band against the number of coordinates $k$ for IRMAE-WD with $\nl=4$ and POD.
For IRMAE-WD, the error in every band except the smallest decreases gradually with $k$ up to the drop position $\kd$, and no band shows an abrupt decrease or stops decreasing earlier than the others.
In other words, the errors in several bands decrease simultaneously in any range of the number of coordinates.
The decrease of the error of IRMAE-WD slows down as $k$ approaches $\kd$ in every band except the smallest, and it stops for $k$ larger than $\kd$.
This is consistent with the result that truncating the model at the drop position gives almost the same accuracy as the full width (section~\ref{subsec:results_warmstart}).
For POD, in contrast, the error in every band except the smallest continues to decrease beyond the same number of coordinates.
Hence, the error of IRMAE-WD stops decreasing at $\kd$ because the model does not use the directions beyond $\kd$ for the reconstruction.
That is, the drop position does not correspond to a threshold of the flow.
In addition, the error in every band of POD except the smallest also decreases gradually with $k$, as for IRMAE-WD.
This suggests that the continuous decrease is independent of the representation and results from the fluctuations of the velocity field being distributed over a wide range of scales.
The added coordinates do not represent a particular scale alone, but contribute a little both to representing the large-scale structures more precisely and to reproducing smaller structures.
From these results, we consider that the reconstruction of each scale improves gradually over many coordinates, that every direction contributes a little to the reconstruction, and that no direction is clearly unnecessary.
Accordingly, the number of coordinates needed to represent the flow should be stated together with the scale to be reproduced, not only with the overall reconstruction rate.

\FloatBarrier
\section{Conclusions}
\label{sec:conclusion}

We applied the implicit rank-minimizing autoencoder with weight decay (IRMAE-WD), based on a three-dimensional multi-scale autoencoder with convolutional neural networks (CNN-AE), directly to the velocity fields of a minimal turbulent channel flow at the friction Reynolds number of $\Retau=110$, and examined the relationship between the singular value spectrum of the latent variables and the intrinsic degrees of freedom of the flow.

We found that a clear drop appeared in the latent spectra of IRMAE-WD and that truncating the models at the drop position lowered the reconstruction rate by at most 0.01 percentage points.
The drop position thus accurately indicates the effective dimension of the trained representation.
However, the drop position moved to the lower-dimension side as linear layers were added, and it kept moving during training in transfer learning.
In addition, the energy of the reconstruction error decreased gradually with the number of coordinates in every wavelength band except the smallest, without an abrupt decrease in any band.
This is probably because, in the minimal channel flow, whose fluctuation energy is spread over a broad range of wavelengths, each added coordinate contributes a little both to representing the large-scale structures more precisely and to reproducing small-scale structures.
Consequently, no direction is clearly unnecessary for the reconstruction.
In this case, the drop position is likely set not only by the flow itself but also by the strength of the implicit regularization, which depends on the number of linear layers.
Thus, for turbulent flows that contain fluctuations over a continuous and wide range of scales, such as the minimal channel flow, the drop position cannot be regarded as the intrinsic degrees of freedom of the flow.

In contrast, for IRMAE-WD obtained by transfer learning, the smallest number of coordinates needed to reach a given accuracy by truncation was almost the same regardless of the number of linear layers, and it was smaller than that of proper orthogonal decomposition (POD) for reconstruction rates up to about 97\% and smaller than that of the baseline CNN-AE, which has no linear layers, up to about 93\%.
The models trained from random initialization with two or three linear layers were also able to reach the same error as POD with up to 17\% fewer coordinates.
Furthermore, when the required number of coordinates was paired with the smallest scale reproduced with a scale-wise reconstruction rate of at least 0.5, IRMAE-WD reproduced a given scale with fewer coordinates than POD.
At reconstruction rates above about 97\%, however, POD reached the same error with fewer coordinates.
The breakdown of the error showed that, at such high accuracy, IRMAE-WD left more error than POD in the large-scale structures, which carry most of the energy, whereas the errors remaining in small-scale structures contributed little to the total error.

From these results, we conclude that IRMAE-WD with truncated latent variables can represent the present minimal turbulent channel flow with fewer coordinates than POD when the required reconstruction rate is up to about 97\% or when the scale to be reproduced is specified.
We also suggest that, when the reconstruction error decreases continuously with the number of coordinates, as in the minimal channel flow, the number of coordinates needed to represent wall turbulence should be stated together with the required accuracy or the scale to be reproduced, rather than as a single number.
Future work includes evaluating the variation of the drop position with additional seeds, applying the method to more complex flows such as separated flows, and using the representation in reduced-order models of the temporal evolution.

\section*{Acknowledgments}
This work was supported through JSPS KAKENHI (Grant No. 26K21736) by Japan Society for the Promotion of Science.

\appendix
\section{Temporal correlation and statistics of the DNS data}
\label{app:dns_correlation}

\begin{figure}[tb]
 \centering
 \includegraphics[width=\linewidth]{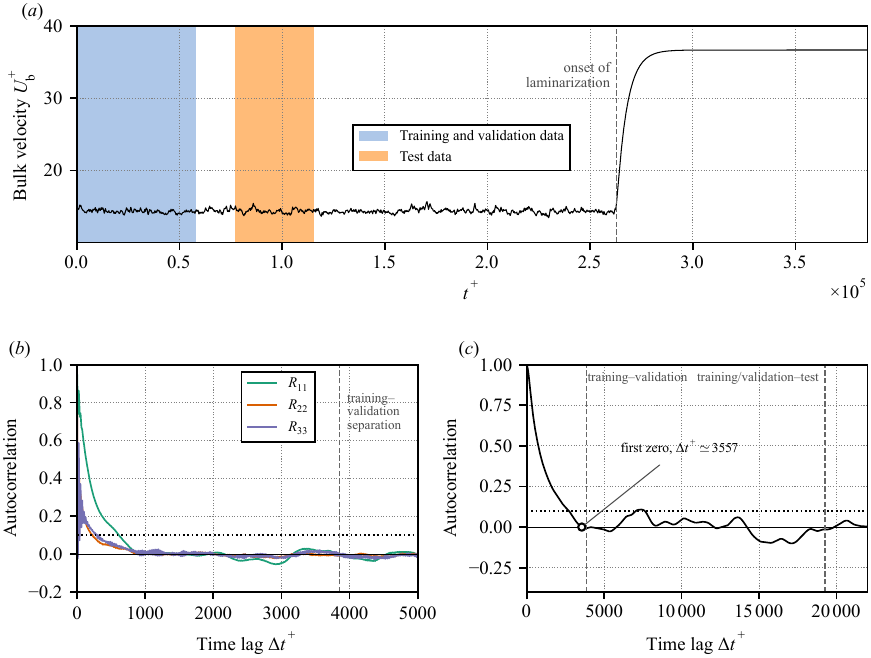}
 \caption{DNS used for the datasets: (\textit{a}) time history of the bulk velocity; (\textit{b}) temporal autocorrelation of the velocity fluctuations at $y^+\simeq15.5$; (\textit{c}) temporal autocorrelation of the bulk velocity before laminarization. Dotted lines: $R=0.1$.}
 \label{fig:app_dns}
\end{figure}

Figure~\ref{fig:app_dns}(\textit{a}) shows the time history of the bulk velocity $U^+_\mathrm{b}$ of the DNS from which the data were obtained.
The bulk velocity $U^+_\mathrm{b}$ increases from $t^+\simeq2.6\times10^5$ and approaches the laminar value ($\Retau/3\simeq36.7$); that is, the flow in this DNS becomes laminar from this time on.
The training, validation, and test data are all taken from the turbulent state before this laminarization.

To check whether the intervals between the datasets are long enough, we examine the temporal correlation of the velocity.
Figure~\ref{fig:app_dns}(\textit{b}) shows the temporal autocorrelations $R_{11}$, $R_{22}$, and $R_{33}$ of $u'$, $v'$, and $w'$ on the plane at $y^+\simeq15.5$, averaged over all grid points in the $x$ and $z$ directions.
The autocorrelations are computed from the 15\,000 snapshots in the period of the training and validation data.
All three autocorrelations reach zero by a time lag of $\Delta t^+\simeq1000$, and for time lags longer than the minimum interval $\Delta t^+\simeq3850$ between the training and validation data, their absolute values are 0.09 or less.
As shown in figure~\ref{fig:app_dns}(\textit{c}), the autocorrelation of the bulk velocity decays more slowly than that of the local velocity and first reaches zero at $\Delta t^+\simeq3560$.
This autocorrelation is computed from the whole history of the bulk velocity before the laminarization.
Both the minimum interval between the training and validation data ($\Delta t^+\simeq3850$) and the interval between these data and the test data ($\Delta t^+\simeq19\,250$) are longer than this time lag.
Hence, the three datasets can be regarded as almost uncorrelated with each other, even in terms of the bulk velocity.

\begin{figure}[tb]
 \centering
 \includegraphics[width=\linewidth]{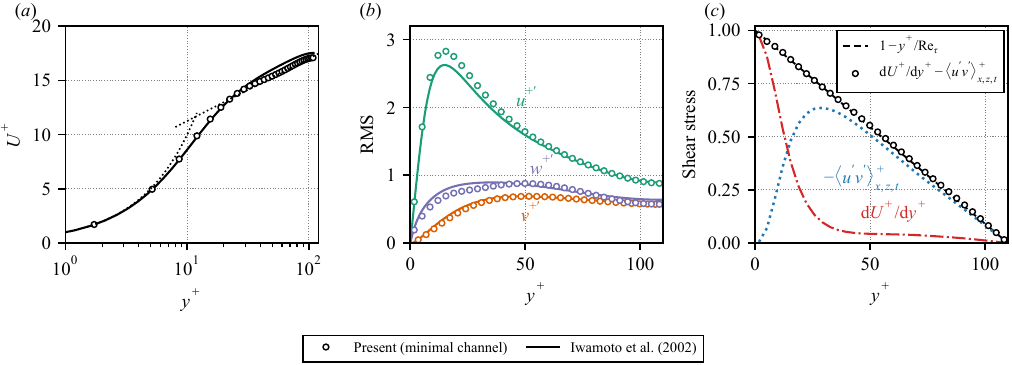}
 \caption{Statistics of the present DNS compared with the reference DNS at $\mathrm{Re}_\tau=109.4$ \citep{Iwamoto2002Database,Iwamoto2002DNS}: (\textit{a}) mean streamwise velocity, with $U^+=y^+$ and $U^+=\ln y^+/0.4+5.5$ (dotted lines); (\textit{b}) root-mean-square velocity fluctuations; (\textit{c}) shear stress balance of the present DNS. Symbols: present DNS, averaged over the whole turbulent period before laminarization (680 instantaneous fields, both walls).}
 \label{fig:app_dns_reference}
\end{figure}

Next, figures~\ref{fig:app_dns_reference}(\textit{a}) and (\textit{b}) compare the wall-normal profiles of the mean velocity $U^+$ and the RMS velocity fluctuations of the present DNS with those of a DNS at $\Retau=109.4$ computed by a spectral method in a large domain ($5\pi\times2\times2\pi$) \citep{Iwamoto2002Database,Iwamoto2002DNS}.
The statistics of the present DNS are computed from 680 instantaneous fields saved over the whole turbulent period before the laminarization ($t^+\leq2.6\times10^5$), by averaging over time and the $x$ and $z$ directions and then over the two walls.
The mean velocity agrees with that of the reference DNS within 1.3\% for $y^+\leq30$, whereas it is up to 4\% smaller in the outer layer away from the wall.
This difference in the outer layer is probably because the larger structures of the outer layer do not fit in the small domain \citep{JimenezMoin1991}.
The peak of the RMS of $u'$ is located at $y^+\simeq15$, as in the reference DNS, but is about 8\% larger.
In addition, the RMS values of $v'$ and $w'$ are smaller than those of the reference DNS by up to 15\% and 18\%, respectively, for $5<y^+\leq30$.
Similarly, for a minimal channel at $\Retau=180$ with the same domain size as the present one and a grid refined near the wall, it has been reported that the mean velocity almost agrees with that in a larger domain, whereas the RMS values deviate, with a larger peak of $u'$ and a smaller $w'$ near the wall \citep{Oura2026PhD}.
These differences are thus likely due mainly to the small domain.
However, because all the methods in the present study are applied to the same DNS data, these differences do not affect the comparison between the methods.
Figure~\ref{fig:app_dns_reference}(\textit{c}) shows the shear stress balance of the present DNS.
In a fully developed channel flow, the sum of the viscous stress $\mathrm{d}U^+/\mathrm{d}y^+$ and the Reynolds shear stress $-\langle u'v'\rangle^+_{x,z,t}$ equals $1-y^+/\Retau$.
In the present DNS, this relation holds within 0.8\% of the wall shear stress.
This indicates that the present DNS is fully developed and that its statistics are converged.

\section{Effect of periodic padding}
\label{app:padding}

\begin{figure}[tb]
 \centering
 \includegraphics[width=\linewidth]{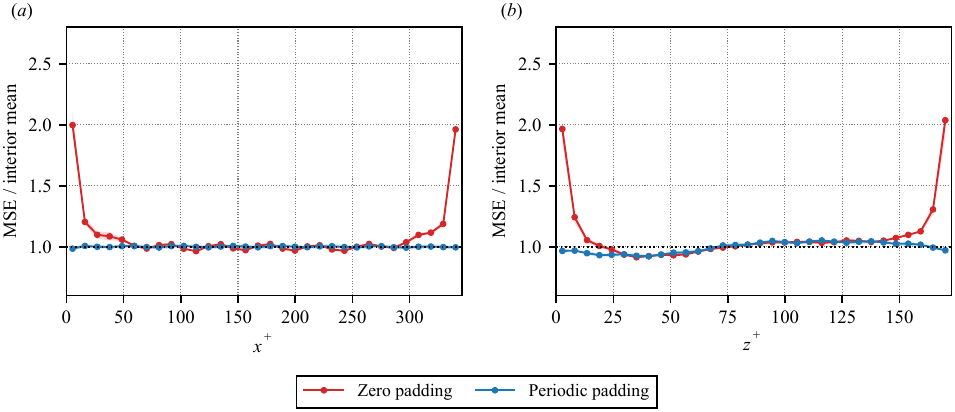}
 \caption{MSE of the baseline CNN-AE ($F=1536$) on the validation data at each grid point in (\textit{a}) $x$ and (\textit{b}) $z$, normalized by the mean over the interior points.}
 \label{fig:app_padding}
\end{figure}

When zeros are padded at the boundaries in the $x$ and $z$ directions in the convolutions of a CNN, the neighboring relation given by the periodic boundary conditions is lost at the grid points near the boundaries.
To examine this effect, we computed the mean squared error on the validation data at each grid point in the $x$ and $z$ directions for the baseline CNN-AE ($F=1536$, three seeds each) trained with zero padding and with periodic padding.
Figure~\ref{fig:app_padding} shows the distributions normalized by the mean over the interior grid points, excluding four grid points at each end.
With zero padding, the errors at the end grid points are 1.9--2.1 times those in the interior, and the errors are large up to about four grid points from the ends ($x^+\simeq43$ and $z^+\simeq22$).
With periodic padding, the errors at the ends are 0.96--1.00 times those in the interior, and the error is almost uniform over the whole domain.
The total error on the validation data is 0.0242 with zero padding and 0.0248 with periodic padding (averages over three seeds), which are comparable within the seed-to-seed variation (0.0211--0.0274 with periodic padding).
Periodic padding thus removes the bias of the error near the boundaries without changing the overall accuracy.

\section{Training stabilization and optimization problems}
\label{app:stabilization}

This appendix describes the optimization problems encountered in applying IRMAE-WD to the three-dimensional turbulent velocity field and the conditions under which the training succeeded.

\subsection{Scope of weight decay and gradient clipping}

In preliminary experiments, applying the AdamW decay to all layers, including the CNN, increased the squared error to about 2.7 times that of the baseline CNN-AE, even with a single linear layer ($\nl=1$).
Table~\ref{tab:app_stabilization} compares combinations of the scope of the decay and the gradient clipping.
These preliminary experiments were performed before the main experiments of this paper with a different data split (10\,000 training snapshots, and the last 1000 or 2000 snapshots of the time series as the test data), for models with $\dz=1536$ and $\nl=1$.
When the decay was applied to all layers (Case 0), the test MSE was 0.0787, compared with 0.0290 for the baseline CNN-AE.
When the decay was limited to the linear layers (Case 1) or was not used (Case 2), the error still increased abruptly in the later stage of training.
Clipping alone (Case 3) improved the error, and the combination of the decay limited to the linear layers with clipping (Case 4) gave the smallest error.
When the Case 4 model was evaluated on the same 1000 snapshots as the baseline CNN-AE, its squared error was 0.02920, only 0.6\% larger than that of the baseline CNN-AE (0.02903).
Based on these results, all the training in the present study uses this combination (equations~\eqref{eq:method_adamw} and \eqref{eq:method_clip}).

\begin{table}[tb]
 \centering
 \caption{Preliminary comparison of the scope of weight decay and gradient clipping for $d_z=1536$ and $n_\ell=1$ (earlier data split).}
 \label{tab:app_stabilization}
 \small
 \begin{tabular}{lllll}
  \toprule
  Case & Weight decay & Clipping & Test MSE & Test snapshots \\
  \midrule
  Baseline CNN-AE & none & none & 0.0290 & 1000 \\
  Case 0 & all layers & none & 0.0787 & 1000 \\
  Case 1 & linear layers & none & 0.0557 & 2000 \\
  Case 2 & none & none & 0.0477 & 2000 \\
  Case 3 & none & $c=1$ & 0.0391 & 2000 \\
  Case 4 & linear layers & $c=1$ & 0.0344 & 2000 \\
  \bottomrule
 \end{tabular}
\end{table}

\subsection{Collapse of wide models and transfer learning}

\begin{figure}[tb]
 \centering
 \includegraphics[width=\linewidth]{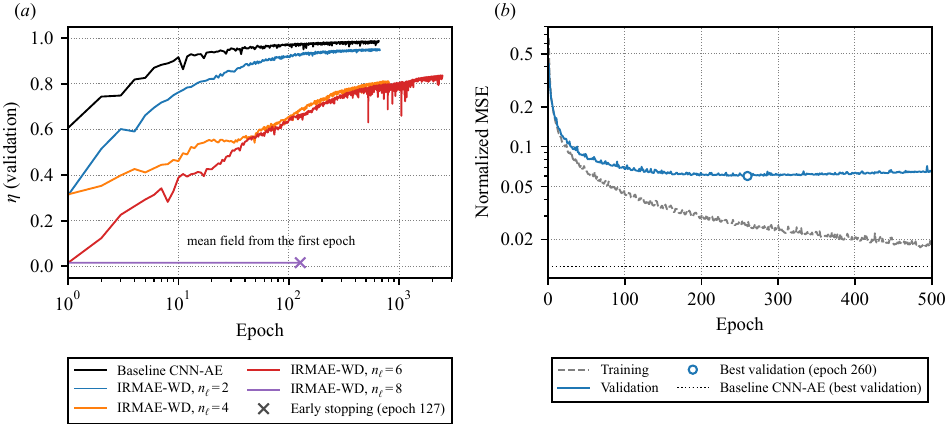}
 \caption{Optimization problems in training from random initialization: (\textit{a}) reconstruction rate on the validation data ($F=3072$, learning rate $10^{-3}$); (\textit{b}) training and validation errors for $n_\ell=2$ with learning rate $10^{-4}$, normalized by the mean squared fluctuation of the validation data.}
 \label{fig:app_pathologies}
\end{figure}

When wide models with $F=3072$ and $F=6144$ were trained from random initialization with the learning rate $10^{-3}$, the baseline CNN-AE reached $\eta=98.8\%$, whereas IRMAE-WD degraded considerably: for $F=3072$, for example, the test reconstruction rate was only 95.3\% even for $\nl=2$.
Figure~\ref{fig:app_pathologies}(\textit{a}) shows the reconstruction rate on the validation data during training for $F=3072$.
The models with more linear layers learned only slowly ($\nl=4$ and 6 with $F=3072$) or fell from the first epoch into a solution with zero fluctuations, i.e., a solution that outputs only the mean field ($\nl=8$ with $F=3072$ and $\nl=4$, 6, and 8 with $F=6144$).
In the models that fell into this solution, the reconstruction rate on the validation data hardly changed after the first epoch, and the training was stopped by early stopping at epoch 127.

When the learning rate was reduced to $10^{-4}$, the model with $F=3072$ and $\nl=2$ did not collapse.
However, as shown in figure~\ref{fig:app_pathologies}(\textit{b}), the validation loss increased after epoch 260 while the training error kept decreasing, and the test reconstruction rate was only 94.2\%.
Reducing the learning rate thus avoided the collapse, but the overfitted model remained far less accurate than the baseline CNN-AE.
In a preliminary experiment, a model with $\dz=1536$ and $\nl=16$ (learning rate $10^{-3}$, decay on the linear layers only, no clipping) showed a similar collapse: the latent variables were strongly biased in one direction, and the decoder fell into the mean-field solution.

This collapse was avoided by transfer learning, in which a trained baseline CNN-AE is used as the initial state, the linear layers are trained first, and then all layers are trained with a small learning rate.
With transfer learning, none of the models with $\nl=2$--10 collapsed, and they reached $\eta=98.0$--98.3\% at the full width.
However, the full-width accuracy of the models obtained by transfer learning did not reach the 98.78\% of the baseline CNN-AE, and the loss of accuracy caused by adding linear layers was not fully recovered even after 3000 epochs (98.11\%, 98.05\%, and 97.95\% for $\nl=6$, 8, and 10).

These results show that, for high-dimensional inputs and latent variables such as three-dimensional turbulent velocity fields, the training of IRMAE-WD can fail even when the baseline CNN-AE is trained successfully.
A failed model can still show a drop: the latent variables of the collapsed model with $\nl=16$ were concentrated in one direction.
Moreover, a model with lower accuracy has its drop position at fewer coordinates (sections~\ref{subsec:results_spectrum} and \ref{subsec:results_warmstart}).
Accordingly, the drop position should be interpreted only for models that have not collapsed and whose accuracy is close to that of the baseline CNN-AE.

\section{Effect of the weight decay coefficient}
\label{app:weight_decay}

To examine the role of weight decay in IRMAE-WD, models with $\dz=1536$ and $\nl=2$ and 4 were trained from random initialization with weight decay coefficients of $\lambda=0$, $10^{-5}$, $10^{-4}$, $10^{-3}$, $10^{-2}$, and $10^{-1}$.
The conditions other than the weight decay coefficient are the same as in section~\ref{subsec:method_training}, and the same initial weights (seed 1) are used for all $\lambda$.
These runs were trained separately from the models in appendix~\ref{app:random_table}.

Figure~\ref{fig:app_wd} shows the singular values of the latent variables of the training data and the drop positions, and table~\ref{tab:app_wd} lists the values for each model.
For both $\nl=2$ and 4, a drop appears even without weight decay ($\lambda=0$).
For $\lambda\leq10^{-2}$, the drop position and the full-width reconstruction rate vary without a systematic trend with $\lambda$, and the variation is much smaller than that caused by changing the number of linear layers.
Only for the largest coefficient, $\lambda=10^{-1}$ with $\nl=4$, does the drop position move to clearly fewer coordinates and the accuracy decrease.
Under the present conditions, weight decay thus changes the drop position and the accuracy substantially only when it is large.
\citet{Zeng2024IRMAEWD} also reported, for the Kuramoto--Sivashinsky equation, that the dimension is estimated correctly over a broad range of the number of linear layers and $\lambda$ and that the estimate fails only when both are large, in agreement with the present results.
A drop without weight decay has also been obtained with linear layers alone in IRMAE \citep{Jing2020IRMAE}.
Together, these results support the interpretation in section~\ref{subsec:results_spectrum} that the drop position is set mainly by the number of linear layers rather than by the weight decay.
Based on these results, we used $\lambda=10^{-3}$ in the other experiments; this value lies in the range ($\lambda\leq10^{-2}$) where the changes in the drop position and the accuracy are much smaller than those caused by changing the number of linear layers.

\begin{figure}[tb]
 \centering
 \includegraphics[width=\linewidth]{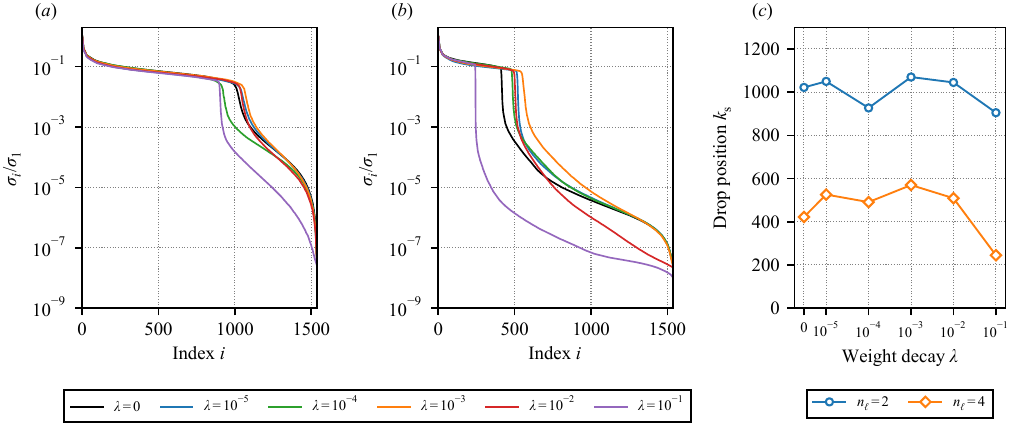}
 \caption{Effect of the weight decay coefficient $\lambda$ for IRMAE-WD trained from random initialization ($d_z=1536$, seed 1): normalized singular values for (\textit{a}) $n_\ell=2$ and (\textit{b}) $n_\ell=4$; (\textit{c}) drop position $k_\mathrm{s}$ against $\lambda$.}
 \label{fig:app_wd}
\end{figure}

\begin{table}[tb]
 \centering
 \caption{IRMAE-WD trained from random initialization with different weight decay coefficients ($d_z=1536$, seed 1).}
 \label{tab:app_wd}
 \small
 \begin{tabular}{rrrrrr}
  \toprule
  $n_\ell$ & $\lambda$ & $k_\mathrm{s}$ & $\Delta_{100}$ & $\eta$ (full) [\%] & $\eta$ ($k_\mathrm{s}$) [\%] \\
  \midrule
    2 & 0 & 1022 & 1.25 & 96.90 & 96.90 \\
  2 & $10^{-5}$ & 1050 & 1.22 & 96.77 & 96.76 \\
  2 & $10^{-4}$ & 927 & 1.36 & 96.18 & 96.18 \\
  2 & $10^{-3}$ & 1070 & 1.29 & 96.55 & 96.54 \\
  2 & $10^{-2}$ & 1045 & 1.45 & 96.82 & 96.82 \\
  2 & $10^{-1}$ & 905 & 2.10 & 96.45 & 96.45 \\
  \midrule
  4 & 0 & 422 & 2.26 & 93.11 & 93.11 \\
  4 & $10^{-5}$ & 526 & 2.49 & 95.01 & 95.01 \\
  4 & $10^{-4}$ & 491 & 2.28 & 94.09 & 94.09 \\
  4 & $10^{-3}$ & 570 & 2.06 & 94.70 & 94.70 \\
  4 & $10^{-2}$ & 509 & 2.47 & 94.33 & 94.33 \\
  4 & $10^{-1}$ & 245 & 3.54 & 90.68 & 90.68 \\
  \bottomrule
 \end{tabular}
\end{table}

\section{Computing environment and training time}
\label{app:environment}

All AEs were trained and evaluated in the same software environment, based on the NVIDIA TensorFlow container (release 25.02), which includes Python 3.12.3, TensorFlow 2.17.0, CUDA 12.8, cuDNN 9.7, and NumPy 1.26.4.
The models were built and trained with the Keras 2 API (tf\_keras).
NVIDIA A100 and NVIDIA RTX PRO 6000 Blackwell Max-Q Workstation Edition GPUs were used, and each model was trained on a single GPU.
In the transfer learning, the linear layers were computed in double precision, and the CNN in single precision.
In the training from random initialization, all layers were computed in single precision.

The random seeds were set separately for the Python standard random number generator, NumPy, and TensorFlow.
We confirmed that the same seed gives the same initial weights in different processes.
However, because deterministic GPU operations were not enforced, the training results may differ slightly between runs even with the same seed.

In training from random initialization ($F=1536$), one epoch took about 30--40~s, and the training until early stopping took, for example, about 10~h for $\dz=1536$ and $\nl=2$ (922 epochs) and about 18~h for $\nl=6$ (2118 epochs).
In step 3 of the transfer learning ($F=3072$, $\nl=4$), one epoch took about 80~s on an A100, and 1000 epochs took about 23~h.

\section{Two definitions of the drop position}
\label{app:knee}

\begin{figure}[tb]
 \centering
 \includegraphics[width=\linewidth]{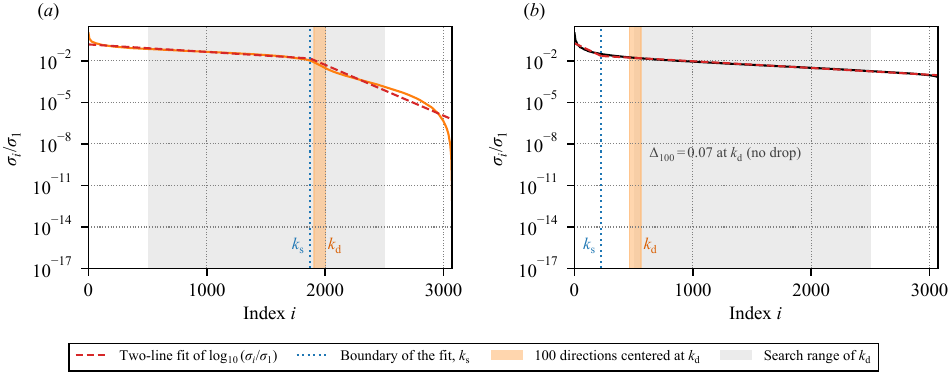}
 \caption{Two definitions of the drop position: (\textit{a}) IRMAE-WD with $n_\ell=4$ obtained by transfer learning; (\textit{b}) baseline CNN-AE ($F=3072$).}
 \label{fig:app_knee}
\end{figure}

In the main text, the drop position $\ks$ of the models trained from random initialization is determined by the two-line fit.
For the 36 models trained from random initialization, $\ks$ obtained from the latent variables of the training data agrees with that obtained from the test data within two indices (exactly for 27 models).
This indicates that the drop position hardly depends on the choice of the data used for the singular value decomposition.
For the models obtained by transfer learning, on the other hand, the spectra fall off steeply at their end.
For the models without a clear drop (the baseline CNN-AE and IRMAE-WD with $\nl=2$), the boundary of the two-line fit is then placed at the first bend of the decrease, near the end of the spectrum, or on a flat part.
As shown in figure~\ref{fig:app_knee}(\textit{b}), for the baseline CNN-AE, which has no drop, the boundary of the two-line fit is placed at the first bend of the decrease ($i=227$), and the largest decrease over 100 directions lies near the lower edge of the search range ($i=514$).
For the models obtained by transfer learning, we instead define the drop position $\kd$ as the center of the largest decrease over 100 directions.
The search range is $512\leq i\leq2500$.
This range excludes such a first bend and the fall-off near the end of the spectra.
When the largest decrease lies at or within a few indices of the edge of the search range, the model is regarded as having no drop.
For models with a clear drop (figure~\ref{fig:app_knee}(\textit{a})), the two definitions differ only slightly: for the weights of $\nl=4$ and 6 at the end of 1000 epochs, $\ks$ of the two-line fit is 1945 and 1874, and $\kd$ is 2053 and 1914, respectively, which differ by 6\% or less.

\FloatBarrier
\section{Results of all models trained from random initialization}
\label{app:random_table}

Table~\ref{tab:app_random} lists, for all 69 IRMAE-WD models trained from random initialization, the drop position $\ks$, the sharpness of the drop $\Delta_{100}$, and the test reconstruction rates at the full width and after truncation at $\ks$.
The 12 configurations with three seeds used in the main text (36 models) are marked with $\ast$.
The remaining 33 models are configurations trained with seed 1 only.
In these configurations as well, $\ks$ and the reconstruction rate at each latent width generally decrease as linear layers are added, although the order of $\nl=4$ and 6 is reversed for some latent widths.
That is, the configurations with seed 1 only show the same tendency as those in the main text (section~\ref{subsec:results_spectrum}).

\begin{table}[!tbp]
 \centering
 \caption{All IRMAE-WD models trained from random initialization ($F=1536$). $k_\mathrm{s}$: drop position of the training latent variables; $\eta$: test reconstruction rate [\%] with the full width and after truncation at $k_\mathrm{s}$. $\ast$: models used in the main text (three seeds). Note: a, reconstruction clearly degraded; b, collapse during training; c, no clear drop ($\Delta_{100}<1$).}
 \label{tab:app_random}
 \footnotesize
 \setlength{\tabcolsep}{3.2pt}
 \begin{tabular}{rrrrrrrl@{\hspace{1.2em}}rrrrrrrl}
  \toprule
  $d_z$ & $n_\ell$ & Seed & $k_\mathrm{s}$ & $\Delta_{100}$ & $\eta$ (full) & $\eta$ ($k_\mathrm{s}$) & Note & $d_z$ & $n_\ell$ & Seed & $k_\mathrm{s}$ & $\Delta_{100}$ & $\eta$ (full) & $\eta$ ($k_\mathrm{s}$) & Note \\
  \midrule
    192 & 2 & 1 & 16 & 0.61 & 89.14 & 61.93 & c & 1024 & 3 & 3$^\ast$ & 565 & 2.14 & 95.30 & 95.30 & \\
  192 & 4 & 1 & 160 & 5.71 & 86.20 & 86.20 & & 1024 & 4 & 1$^\ast$ & 384 & 2.28 & 90.48 & 90.48 & \\
  192 & 6 & 1 & 142 & 5.94 & 86.78 & 86.78 & & 1024 & 4 & 2$^\ast$ & 464 & 2.61 & 94.44 & 94.44 & \\
  192 & 8 & 1 & 135 & 5.04 & 85.67 & 85.67 & & 1024 & 4 & 3$^\ast$ & 480 & 2.58 & 94.21 & 94.20 & \\
  384 & 2 & 1 & 375 & 2.83 & 93.55 & 93.53 & & 1024 & 6 & 1$^\ast$ & 391 & 2.90 & 93.46 & 93.45 & \\
  384 & 4 & 1 & 316 & 4.28 & 92.61 & 92.61 & & 1024 & 6 & 2$^\ast$ & 198 & 3.23 & 86.20 & 86.20 & \\
  384 & 6 & 1 & 280 & 3.99 & 91.79 & 91.79 & & 1024 & 6 & 3$^\ast$ & 208 & 3.19 & 87.03 & 87.03 & \\
  384 & 8 & 1 & 131 & 3.65 & 84.23 & 84.22 & & 1024 & 8 & 1 & 49 & 4.39 & 75.84 & 75.84 & a \\
  512 & 2 & 1 & 507 & 2.13 & 95.44 & 95.44 & & 1152 & 2 & 1 & 878 & 1.57 & 96.13 & 96.13 & \\
  512 & 3 & 1 & 451 & 4.02 & 94.52 & 94.52 & & 1152 & 3 & 1 & 643 & 2.16 & 95.73 & 95.73 & \\
  512 & 4 & 1 & 310 & 2.94 & 92.05 & 92.05 & & 1152 & 4 & 1 & 378 & 2.01 & 90.16 & 90.16 & \\
  512 & 6 & 1 & 295 & 3.12 & 91.72 & 91.71 & & 1280 & 2 & 1$^\ast$ & 971 & 1.47 & 97.17 & 97.16 & \\
  512 & 8 & 1 & 110 & 3.74 & 82.95 & 82.95 & & 1280 & 2 & 2$^\ast$ & 900 & 1.48 & 96.91 & 96.91 & \\
  640 & 2 & 1 & 635 & 2.05 & 96.05 & 96.05 & & 1280 & 2 & 3$^\ast$ & 976 & 1.32 & 96.81 & 96.80 & \\
  640 & 3 & 1 & 536 & 3.00 & 95.37 & 95.37 & & 1280 & 3 & 1$^\ast$ & 650 & 2.03 & 95.71 & 95.70 & \\
  640 & 4 & 1 & 347 & 2.76 & 92.62 & 92.62 & & 1280 & 3 & 2$^\ast$ & 688 & 2.14 & 96.18 & 96.18 & \\
  768 & 2 & 1$^\ast$ & 724 & 3.31 & 96.37 & 96.37 & & 1280 & 3 & 3$^\ast$ & 635 & 2.18 & 95.83 & 95.83 & \\
  768 & 2 & 2$^\ast$ & 707 & 2.42 & 96.35 & 96.34 & & 1280 & 4 & 1 & 364 & 2.07 & 89.36 & 89.36 & \\
  768 & 2 & 3$^\ast$ & 723 & 5.06 & 96.71 & 96.71 & & 1280 & 6 & 1 & 356 & 3.06 & 91.79 & 91.79 & \\
  768 & 3 & 1$^\ast$ & 575 & 2.56 & 95.62 & 95.62 & & 1280 & 8 & 1 & 33 & 4.45 & 70.16 & 70.16 & b \\
  768 & 3 & 2$^\ast$ & 547 & 2.27 & 95.41 & 95.40 & & 1408 & 2 & 1 & 980 & 1.39 & 97.23 & 97.22 & \\
  768 & 3 & 3$^\ast$ & 559 & 2.44 & 95.16 & 95.16 & & 1408 & 3 & 1 & 786 & 1.90 & 96.15 & 96.15 & \\
  768 & 4 & 1$^\ast$ & 491 & 2.85 & 94.79 & 94.78 & & 1408 & 4 & 1 & 449 & 2.43 & 93.97 & 93.97 & \\
  768 & 4 & 2$^\ast$ & 459 & 2.82 & 94.50 & 94.50 & & 1536 & 2 & 1$^\ast$ & 1030 & 1.29 & 97.11 & 97.11 & \\
  768 & 4 & 3$^\ast$ & 517 & 2.83 & 94.87 & 94.87 & & 1536 & 2 & 2$^\ast$ & 991 & 1.31 & 97.13 & 97.12 & \\
  768 & 6 & 1 & 348 & 2.90 & 92.73 & 92.73 & & 1536 & 2 & 3$^\ast$ & 1110 & 1.32 & 97.01 & 97.01 & \\
  768 & 8 & 1 & 73 & 3.74 & 79.47 & 79.47 & a & 1536 & 3 & 1$^\ast$ & 702 & 2.01 & 96.32 & 96.31 & \\
  896 & 2 & 1 & 841 & 2.71 & 97.02 & 97.02 & & 1536 & 3 & 2$^\ast$ & 580 & 1.98 & 94.92 & 94.92 & \\
  896 & 3 & 1 & 623 & 2.30 & 95.67 & 95.66 & & 1536 & 3 & 3$^\ast$ & 539 & 1.94 & 94.51 & 94.51 & \\
  896 & 4 & 1 & 479 & 2.70 & 93.84 & 93.84 & & 1536 & 4 & 1$^\ast$ & 528 & 2.03 & 92.74 & 92.74 & \\
  1024 & 2 & 1$^\ast$ & 916 & 1.92 & 97.20 & 97.19 & & 1536 & 4 & 2$^\ast$ & 418 & 2.60 & 93.69 & 93.69 & \\
  1024 & 2 & 2$^\ast$ & 926 & 1.98 & 97.35 & 97.34 & & 1536 & 4 & 3$^\ast$ & 505 & 2.14 & 92.91 & 92.91 & \\
  1024 & 2 & 3$^\ast$ & 830 & 1.64 & 96.86 & 96.86 & & 1536 & 6 & 1 & 232 & 3.21 & 87.42 & 87.42 & \\
  1024 & 3 & 1$^\ast$ & 648 & 2.23 & 96.10 & 96.09 & & 1536 & 8 & 1 & 8 & 4.54 & 37.52 & 37.52 & a \\
  1024 & 3 & 2$^\ast$ & 602 & 2.28 & 95.88 & 95.88 & & & & & & & & & \\
  \bottomrule
 \end{tabular}
\end{table}

\section{Turbulence statistics of the reconstructed fields}
\label{app:turb_stats}

In this appendix, we compare, in addition to the RMS values shown in section~\ref{subsec:results_physics}, the turbulent kinetic energy and the Reynolds shear stress for IRMAE-WD with $\nl=4$ obtained by transfer learning and for POD.
The models and the numbers of coordinates are the same as in figure~\ref{fig:results_rms}.
For both the reconstructed fields and the DNS, the statistics are computed from the fluctuations $u'$, $v'$, and $w'$ about the profiles averaged over time and the $x$ and $z$ directions of the test data.
The turbulent kinetic energy is $k^+_\mathrm{t}=\frac{1}{2}\langle u'^2+v'^2+w'^2\rangle_{x,z,t}$, and the Reynolds shear stress is $-\langle u'v'\rangle^+_{x,z,t}$; both are functions of $y$.
Below, each quantity is compared by the ratio of its sum over $y$ to that of the DNS.

\begin{figure}[tb]
 \centering
 \includegraphics[width=\linewidth]{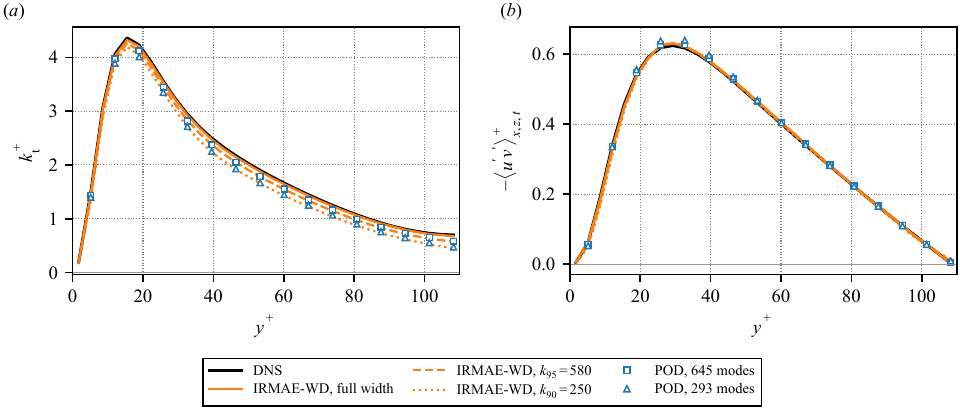}
 \caption{Turbulence statistics on the test data for IRMAE-WD with $n_\ell=4$ obtained by transfer learning and for POD with the same test error as IRMAE-WD at $k_{95}$ and $k_{90}$: (\textit{a}) turbulent kinetic energy; (\textit{b}) Reynolds shear stress.}
 \label{fig:app_turb_stats}
\end{figure}

Figure~\ref{fig:app_turb_stats} shows the results.
The turbulent kinetic energy decreases at all wall-normal locations as the number of coordinates is reduced, and its ratio to the DNS is 0.99 at the full width, 0.95 at $k_{95}$, and 0.89 at $k_{90}$.
The ratios for POD at the same reconstruction rates are almost the same, 0.95 and 0.90.
At $k_{90}$, the ratio is also 0.89 for the baseline CNN-AE and for IRMAE-WD with $\nl=2$--10.
The Reynolds shear stress, in contrast, hardly changes: its ratio to the DNS is 0.99 even at $k_{90}$, and 1.02 for POD.

\FloatBarrier
\bibliographystyle{jfm_mod}
\bibliography{references}
\end{document}